\documentclass[a4paper,12pt]{article}
\usepackage{amsmath}
\usepackage{amssymb}
\usepackage{amsfonts}
\usepackage{amssymb,amsmath}
\usepackage{cancel}
\usepackage[dvips]{lscape,graphicx,epsfig}
\usepackage{fleqn}
\usepackage[footnotesize]{caption}
\usepackage{mathrsfs}

\def\beq{\begin{equation}}
\def\eeq{\end{equation}}
\def\bea{\begin{eqnarray}}
\def\eea{\end{eqnarray}}

\def\<{\left\langle}
\def\>{\right\rangle}

\renewcommand{\baselinestretch}{1.30}

\newcommand{\bc}{\begin{center}}
\newcommand{\ec}{\end{center}}
\newcommand{\bd}{\begin{displaymath}}
\newcommand{\ed}{\end{displaymath}}
\newcommand{\be}{\begin{equation}}
\newcommand{\ee}{\end{equation}}
\newcommand{\ba}{\begin{array}}
\newcommand{\ea}{\end{array}}
\newcommand{\bt}{\begin{tabular}}
\newcommand{\et}{\end{tabular}}

\newcommand{\ds}{\displaystyle}

\begin{document}

\bibliographystyle{OurBibTeX}

\begin{titlepage}

\vspace*{-15mm}
\begin{flushright}
{UH511-1159-2011}\\
\end{flushright}
\vspace*{5mm}

\begin{center}
{
\sffamily
\Large Novel Higgs Decays and Dark Matter in the E$_6$SSM}
\\[8mm]
J.~P.~Hall$^a$\footnote{E-mail: \texttt{jonathan.hall@soton.ac.uk}.},
S.~F.~King$^a$\footnote{E-mail: \texttt{s.f.king@soton.ac.uk}.},
R.~Nevzorov$^b$\footnote{On leave of absence from the Theory Department, ITEP, Moscow, Russia;\\[-2mm]
$~~~~~$ E-mail: \texttt{nevzorov@phys.hawaii.edu}.},
S.~Pakvasa$^b$\footnote{E-mail: \texttt{pakvasa@phys.hawaii.edu}.},
M.~Sher$^c$\footnote{E-mail: \texttt{mtsher@wm.edu}.}
\\[3mm]
{\small\it
$^a$ School of Physics and Astronomy, University of Southampton,\\
Southampton, UK  \\[2mm]
$^b$ Department of Physics and Astronomy, University of Hawaii,\\
Honolulu, USA \\[2mm]
$^c$ Physics Department, College of William and Mary,\\
Williamsburg, USA}\\[2mm]
\end{center}
\vspace*{0.75cm}

\begin{abstract}
\noindent
The Exceptional Supersymmetric (SUSY) Standard Model (E$_6$SSM)
predicts three families of Higgs doublets plus three Higgs singlets, where
one family develops vacuum expectation values (VEVs), while the remaining
two which do not are called Inert. The model can account for the dark matter
relic abundance if the two lightest Inert neutralinos, identified as the
(next-to) lightest SUSY particles ((N)LSPs), have masses close to half the
$Z$ mass. In this case we find that the usual SM-like Higgs boson
decays more than 95\% of the time into either LSPs or NLSPs.
The latter case produces a final state containing two leptons $l^{+} l^{-}$
with an invariant mass less than or about 10 GeV. We illustrate this scenario
with a set of benchmark points satisfying phenomenological constraints and
the WMAP dark matter relic abundance. This scenario also predicts other
light Inert chargino and neutralino states below $200\,\mbox{GeV}$, and
large LSP direct detection cross-sections close to current limits and
observable soon at XENON100.
\end{abstract}

\end{titlepage}
\newpage
\setcounter{footnote}{0}

\section{Introduction}

The discovery of the Higgs boson, the last missing piece in the
Standard Model (SM) of particle physics, is one of the main goals
of upcoming accelerators. It is expected that the Higgs particle will
be detected at the Large Hadron Collider (LHC) in the near future.
The strategy for Higgs searches depends on the decay branching fractions
of the Higgs boson to different channels. Physics beyond the Standard
Model may affect the Higgs decay rates to SM particles and give
rise to new channels of Higgs decays requiring a drastic change in
the strategy for Higgs boson searches (for recent reviews of
nonstandard Higgs boson decays see \cite{Chang:2008cw}). In particular,
there exist several extensions of the Standard Model in which the
Higgs boson can decay with a substantial branching fraction into
particles which can not be directly detected. Invisible Higgs decay
modes may occur in models with an enlarged symmetry breaking sector
(Majoron models, SM with extra singlet scalar fields etc.)
\cite{majoron}-\cite{Martin:1999qf}, in ``hidden valley'' models \cite{hidden-valley},
in the SM with a fourth generation of fermions
\cite{fourth-generation}-\cite{diffraction}, in the models with compact
and large extra dimensions \cite{Martin:1999qf},
\cite{higgs-extraD}-\cite{Battaglia:2004js},
in the littlest Higgs model with T-parity \cite{Asano:2006nr} etc.
\footnote{In the context of the so-called ``nightmare scenario'',
in which the LHC produces nothing beyond the SM, Higgs decays
into and interactions with dark matter particles were discussed in \cite{Kanemura:2010sh}.}.

Another example wherein invisible decay modes can occur
is supersymmetry (SUSY), with the lightest Higgs boson
decaying into the lightest SUSY particle (LSP). R--parity conservation
ensures the stability of the LSP so that the LSP can play the role of
cold dark matter (CDM) \cite{Bertone:2004pz}. In most scenarios the LSP
is the lightest neutralino, a linear combination of neutral electroweak
(EW) gauginos and Higgsinos.  In some regions
of the parameter space the lightest Higgs boson in the Minimal Supersymmetric
Standard Model (MSSM) decays into the lightest neutralino with a
relatively large branching ratio, therefore giving rise to invisible
final states if R--parity is conserved \cite{Baer:1987eb}.

LEP and Tevatron data still allow the neutralino LSP to be sufficiently light that the decays of the lightest Higgs into these neutralinos is kinematically allowed.    Light neutralinos can annihilate efficiently
through a Z-pole resulting in a reasonable density of dark matter.
Moreover the Cryogenic Dark Matter Search (CDMS) experiment recently
reported the observation of 2 events possibly due to dark matter
scattering with an expected background of about 0.8 events \cite{Ahmed:2009zw}.
CDMS events suggest that
the mass of the dark matter particles are around $40-80\,\mbox{GeV}$
while their spin independent elastic cross section is
$\sigma_{SI}\approx \mbox{few}\times 10^{-44}\,\mbox{cm}^2$.
If recent results of the CDMS experiment get confirmed then
scenarios with invisible decays of the Higgs boson will
become rather plausible.

Certainly the presence of invisible decays modifies considerably
Higgs boson searches, making Higgs discovery much more difficult. If
the Higgs is mainly invisible, then the visible  branching ratios will be
dramatically reduced, preventing detection in the much studied
channels at the LHC and the Tevatron. In the case where invisible
Higgs decays dominate it is impossible to fully reconstruct
a resonance and it is very challenging to identify it at the collider
experiments, i.e. quantum numbers remain unknown. At $e^{+} e^{-}$
colliders, the problems related to the observation of the invisible
Higgs are less severe \cite{Eboli:1994bm},\cite{Djouadi:1994mr} since it
can be tagged through the recoiling $Z$.
Presently, the LEP II collaborations exclude invisible Higgs masses
up to 114.4 GeV \cite{:2001xz}\footnote{Similar limits could apply to 
the case where the Higgs decays some fraction of the time into soft 
lepton pairs plus missing energy, as will be the case for some of 
the novel Higgs decays considered in this paper.}.

On the other hand,  Higgs searches at hadron colliders are
more difficult in the presence of such invisible decays. Previous
studies have analysed $ZH$ and $WH$ associated
production \cite{Choudhury:1993hv}, \cite{Godbole:2003it}-\cite{Davoudiasl:2004aj}
as well as $t\bar{t} H$ production \cite{Gunion:1993jf} and $t\bar{t} VV$ 
($b\bar{b} VV$) production \cite{Boos:2010pu} as promising channels.
The possibility of observing an ``invisible'' Higgs boson in central
exclusive diffractive production at the LHC was studied in \cite{diffraction}.
Another proposal is to observe such an invisible Higgs in inelastic events
with large missing transverse energy and two high $E_T$ jets. In this case
the Higgs boson is produced by $VV$ fusion and has large transverse momentum
resulting in a signal with two quark jets with distinctive
kinematic distributions as compared to $Zjj$ and $Wjj$ backgrounds
\cite{Battaglia:2004js},\cite{Davoudiasl:2004aj},\cite{Eboli:2000ze}.

Consideration of the possibility that the dominant Higgs decays will be invisible
would lead to drastic changes
in the strategy of Higgs boson searches. Therefore, it is  a rather
interesting subject of investigation as to the nature and extent of
invisibility acquired by Higgs, and how it can be related to specific
aspects of the models concerned, especially well motivated SUSY
extensions of the SM. In this article, we consider the exotic decays
of the lightest Higgs boson and associated novel collider signatures
within the Exceptional Supersymmetric Standard Model (E$_6$SSM)
\cite{King:2005jy}-\cite{King:2005my}. This $E_6$ inspired SUSY model
is based on the low--energy Standard Model gauge group together with
an extra $U(1)_{N}$ gauge symmetry under which right-handed neutrinos
have zero charge. In the E$_6$SSM the $\mu$ problem is solved in a
similar way as in the Next--to--Minimal Supersymmetric Standard Model
(NMSSM), but without the accompanying problems of
singlet tadpoles or domain walls. Because right--handed neutrinos do
not participate in the gauge interactions in this model they
can be superheavy, shedding light on the origin of the mass hierarchy
in the lepton sector and providing a mechanism for the generation of
the baryon asymmetry in the Universe via leptogenesis \cite{King:2008qb}.

The particle spectrum of the E$_6$SSM contains exotic matter.
In particular, it involves three SM singlet superfields that have
non-zero $U(1)_{N}$ charges. One of these singlets acquires a non-zero
vacuum expectation value (VEV), breaking $U(1)_{N}$ symmetry and
inducing the effective $\mu$ term and the masses of the the exotic fermions.
The masses of the fermion components of the two other singlet
superfields (Inert neutralinos) are also related to the VEVs of the Higgs
doublets. Because the Yukawa couplings that determine the strength
of these interactions are constrained by the requirement of the
validity of perturbation theory up to the Grand Unification scale
the masses of the corresponding Inert neutralinos are expected to
be lighter than $60-65\,\mbox{GeV}$. As a result the lightest Inert
neutralino tends to be the lightest SUSY particle in the spectrum.
Such a neutralino can give an appropriate contribution to the dark
matter density consistent with the recent observations if it has
mass $35-50\,\mbox{GeV}$ \cite{Hall:2009aj}. In this case the lightest Higgs
boson decays predominantly into Inert neutralino whereas usual Higgs
branching ratios are less than a few percent.

The layout of this paper is as follows. In Section 2
we briefly review the E$_6$SSM. In Sections 3 and 4 the spectrum and
couplings of the Inert neutralinos, charginos and Higgs bosons are
specified. The novel decays of the lightest CP-even Higgs state and dark matter constraints are
discussed in Section 5. In section 6 we discuss the benchmark points and the experimental constraints
and predictions. Section 7 summarizes and concludes the paper.

\section{Exceptional SUSY model}

\noindent
The E$_6$SSM is based on the $SU(3)_C\times SU(2)_W\times U(1)_Y \times U(1)_N$
gauge group which is a subgroup of $E_6$. The additional low energy $U(1)_N$,
which is not present either in the SM or in the MSSM, is a linear superposition
of $U(1)_{\chi}$ and $U(1)_{\psi}$, namely
\be
U(1)_N=\dfrac{1}{4} U(1)_{\chi}+\dfrac{\sqrt{15}}{4} U(1)_{\psi}\,,
\label{lg1}
\ee
where the $U(1)_{\psi}$ and $U(1)_{\chi}$ symmetries are defined by:
$$
E_6\to SO(10)\times U(1)_{\psi}\,,\qquad SO(10)\to SU(5)\times U(1)_{\chi}\,.
$$
Thus the E$_6$SSM can originate from an $E_6$ GUT gauge group which is broken at the
GUT scale $M_X$. The extra $U(1)_N$ gauge symmetry is defined such that right--handed
neutrinos carry zero charges.

In $E_6$ theories the anomalies cancel automatically; all models that are
based on the $E_6$ subgroups and contain complete representations of $E_6$ should
be anomaly--free. Consequently, in order to make a supersymmetric model with an
extra $U(1)_{N}$ anomaly--free, one is forced to augment the minimal particle spectrum
by a number of exotics which, together with ordinary quarks and leptons, form
complete fundamental $27$ representations of $E_6$. Thus the particle content of the
E$_6$SSM involves at least three fundamental representations of $E_6$ at low energies.
These multiplets decompose under the $SU(5)\times U(1)_{N}$ subgroup of $E_6$
as follows \cite{201}:
\begin{equation}
\begin{array}{rcl}
27_i&\to &\ds\left(10,\,\dfrac{1}{\sqrt{40}}\right)_i+\left(5^{*},\,\dfrac{2}{\sqrt{40}}\right)_i
+\left(5^{*},\,-\dfrac{3}{\sqrt{40}}\right)_i +\ds\left(5,-\dfrac{2}{\sqrt{40}}\right)_i\\[2mm]
&&+\left(1,\dfrac{5}{\sqrt{40}}\right)_i+\left(1,0\right)_i\,.
\end{array}
\label{essm1}
\end{equation}
The first and second quantities in brackets are the $SU(5)$ representation and
extra $U(1)_{N}$ charge respectively, while $i$ is a family index that runs from 1 to 3.
An ordinary SM family, which contains the doublets of left--handed quarks $Q_i$ and
leptons $L_i$, right-handed up-- and down--quarks ($u^c_i$ and $d^c_i$) as well as
right--handed charged leptons, is assigned to
$\left(10,\,\dfrac{1}{\sqrt{40}}\right)_i$ + $\left(5^{*},\,\dfrac{2}{\sqrt{40}}\right)_i$.
Right-handed neutrinos $N^c_i$ should be associated with the last term in Eq.~(\ref{essm1}),
$\left(1,\, 0\right)_i$. Because they do not carry any charges right-handed neutrinos are
expected to be superheavy allowing them to be used for both the see--saw mechanism and
leptogenesis. The next-to-last term, $\left(1,\, \dfrac{5}{\sqrt{40}}\right)_i$, represents
SM-singlet fields $S_i$, which carry non-zero $U(1)_{N}$ charges and therefore survive
down to the EW scale. The pair of $SU(2)_W$--doublets ($H^d_{i}$ and $H^u_{i}$) that are
contained in $\left(5^{*},\,-\dfrac{3}{\sqrt{40}}\right)_i$
and $\left(5,\,-\dfrac{2}{\sqrt{40}}\right)_i$ have the quantum numbers of Higgs doublets.
They form either Higgs or Inert Higgs $SU(2)_W$ multiplets~\footnote{We use the terminology
``Inert Higgs'' to denote Higgs--like doublets that do not develop VEVs.}.  Other components
of these $SU(5)$ multiplets form colour triplets of exotic quarks $\overline{D}_i$ and $D_i$
with electric charges $-1/3$ and $+1/3$, respectively. These exotic quark states carry a $B-L$
charge $\pm2/3$, twice that of ordinary ones. Therefore in
phenomenologically viable $E_6$ inspired models they can be either diquarks or leptoquarks.

In addition to the complete $27_i$ multiplets the low energy matter content of the E$_6$SSM
can be supplemented by an $SU(2)_W$ doublet $\hat{L}_4$ and anti-doublet $\hat{\overline{L}}_4$
from the extra $27'$ and $\overline{27'}$ to preserve gauge coupling unification. These
components of the $E_6$ fundamental representation originate from
$\left(5^{*},\,\dfrac{2}{\sqrt{40}} \right)$ of $27'$ and $\left(5,\,-\dfrac{2}{\sqrt{40}}\right)$
of $\overline{27'}$ by construction.

Thus, in addition to a $Z'$ corresponding to the
$U(1)_N$ symmetry, the E$_6$SSM involves extra matter beyond the MSSM that forms three $5+5^{*}$
representations of $SU(5)$ plus three $SU(5)$ singlets with $U(1)_N$ charges. The analysis performed
in \cite{King:2007uj} shows that the unification of gauge couplings in the E$_6$SSM can be achieved
for any phenomenologically acceptable value of $\alpha_3(M_Z)$ consistent with the measured low
energy central value, unlike in the MSSM which, ignoring the effects of high energy threshold
corrections, requires values of $\alpha_3(M_Z)$ which are significantly above the experimentally
measured central value. The presence of a $Z'$ boson and of exotic quarks predicted by the E$_6$SSM
provides spectacular new physics signals at the LHC which were discussed in
\cite{King:2005jy}--\cite{King:2005my}, \cite{Accomando:2006ga}. Recently the particle spectrum and
collider signatures associated with it were studied within the constrained version of the
E$_6$SSM \cite{202}.

In general, the $E_6$ symmetry does not forbid lepton and baryon number violating operators that result
in rapid proton decay. Moreover, exotic particles in $E_6$ inspired SUSY models give rise to new
Yukawa interactions that induce unacceptably large non--diagonal flavour transitions. To suppress
these effects in the E$_6$SSM an approximate $Z^{H}_2$ symmetry is imposed. Under this symmetry
all superfields except one pair of $H^d_{i}$ and $H^u_{i}$ (say $H_d\equiv H^d_{3}$ and
$H_u\equiv H^u_{3}$) and one SM-type singlet field ($S\equiv S_3$) are odd. The $Z^{H}_2$ symmetry
reduces the structure of the Yukawa interactions to
\begin{eqnarray}
W_{\rm E_6SSM}&\simeq &  \lambda \hat{S} (\hat{H}_u \hat{H}_d)+
\lambda_{\alpha\beta} \hat{S} (\hat{H}^d_{\alpha} \hat{H}^u_{\beta})
+\tilde{f}_{\alpha\beta} \hat{S}_{\alpha} (\hat{H}^d_{\beta}\hat{H}_u)
+f_{\alpha\beta} \hat{S}_{\alpha} (\hat{H}_d \hat{H}^u_{\beta})
+\kappa_{ij} \hat{S} (\hat{D}_i\hat{\overline{D}}_j)
\nonumber\\[2mm]
&+&
h^U_{ij}(\hat{H}_{u} \hat{Q}_i)\hat{u}^c_{j} + h^D_{ij}(\hat{H}_{d} \hat{Q}_i)\hat{d}^c_j
+ h^E_{ij}(\hat{H}_{d} \hat{L}_i)\hat{e}^c_{j}+ h_{ij}^N(\hat{H}_{u} \hat{L}_i)\hat{N}_j^c\nonumber\\[2mm]
&+& \dfrac{1}{2}M_{ij}\hat{N}^c_i\hat{N}^c_j+\mu'(\hat{L}_4\hat{\overline{L}}_4)+
h^{E}_{4j}(\hat{H}_d \hat{L}_4)\hat{e}^c_j
+h_{4j}^N(\hat{H}_{u}\hat{L}_4)\hat{N}_j^c\,,
\label{essm2}
\end{eqnarray}
where $\alpha,\beta=1,2$ and $i,j=1,2,3$\,. The $SU(2)_W$ doublets $\hat{H}_u$ and $\hat{H}_d$
and SM-type singlet field $\hat{S}$, that are even under the $Z^{H}_2$ symmetry,
play the role of Higgs fields. At the physical vacuum the Higgs fields develop VEVs
\be
\langle H_d\rangle =\ds\frac{1}{\sqrt{2}}\left(
\begin{array}{c}
v_1\\ 0
\end{array}
\right) , \qquad
\langle H_u\rangle =\ds\frac{1}{\sqrt{2}}\left(
\begin{array}{c}
0\\ v_2
\end{array}
\right) ,\qquad
\langle S\rangle =\ds\frac{s}{\sqrt{2}}.
\label{41}
\ee
generating the masses of the quarks and leptons. Instead of $v_1$ and $v_2$ it is more convenient to use
$\tan\beta=v_2/v_1$ and $v=\sqrt{v_1^2+v_2^2}=246\,\mbox{GeV}$. The VEV of the SM-type singlet field, $s$,
breaks the extra $U(1)_N$ symmetry thereby providing an effective $\mu$ term as well as the necessary
exotic fermion masses and also inducing that of the $Z'$ boson. Therefore the singlet field $S$ must acquire
a large VEV in order to avoid conflict with direct particle searches at present and past accelerators.
This also requires the Yukawa couplings $\lambda_i$ and $\kappa_i$ to be reasonably large. If $\lambda_i$
or $\kappa_i$ are large enough at the GUT scale they affect the evolution of the soft scalar mass $m_S^2$
of the singlet field $S$ rather strongly resulting in a negative value of $m_S^2$ at low energies which
triggers the breakdown of the $U(1)_{N}$ symmetry.

Note that the surviving components from the $27'$ and $\overline{27'}$ manifest themselves
in the Yukawa interactions (\ref{essm2}) as fields with lepton number $L=\pm 1$. The corresponding
mass term $\mu'L_4\overline{L}_4$ in the superpotential (\ref{essm2}) is not involved in the process
of electroweak symmetry breaking (EWSB). Moreover this term is not suppressed by the $E_6$ symmetry.
Therefore the parameter $\mu'$ remains arbitrary. Gauge coupling unification requires $\mu'$ to be below about
$100\,\mbox{TeV}$ \cite{King:2007uj}. Thus we assume that the scalar and fermion components of the
superfields $\hat{L}_4$ and $\hat{\overline{L}_4}$ are very heavy so that they decouple from the
rest of the particle spectrum.

Although $Z^{H}_2$ eliminates any problems related with baryon number violation and non-diagonal
flavour transitions it also forbids all Yukawa interactions that would allow the exotic quarks
to decay. Since models with stable charged exotic particles are ruled out by various
experiments \cite{203} the $Z^{H}_2$ symmetry must be broken. At the same time, the breakdown of
$Z^{H}_2$ should not give rise to operators that would lead to rapid proton decay. There are two ways
to overcome this problem: the Lagrangian must be invariant with respect to either a $Z_2^L$ symmetry,
under which all superfields except leptons are even (Model I), or a $Z_2^B$ discrete symmetry,
which implies that exotic quark and lepton superfields are odd whereas the others remain even
(Model II). If the Lagrangian is invariant under the $Z_2^L$ symmetry, then the
terms in the superpotential which permit exotic quarks to decay and are allowed by the $E_6$ symmetry
can be written in the form
\begin{equation}
W_1=g^Q_{ijk}\hat{D}_{i} (\hat{Q}_j \hat{Q}_k)+
g^{q}_{ijk}\hat{\overline{D}}_i \hat{d}^c_j \hat{u}^c_k\,,
\label{essm3}
\end{equation}
that implies that exotic quarks are diquarks. If $Z_2^B$ is imposed then the following
couplings are allowed:
\begin{equation}
W_2=g^E_{ijk} \hat{e}^c_i \hat{D}_j \hat{u}^c_k+
g^D_{ijk} (\hat{Q}_i \hat{L}_j) \hat{\overline{D}}_k\,.
\label{essm4}
\end{equation}
In this case baryon number conservation requires the exotic quarks to be leptoquarks.

\section{Inert charginos and neutralinos}

From here on we assume that $Z^{H}_2$ symmetry violating couplings are small
and can be neglected in our analysis. This assumption can be justified if
we take into account that the $Z^{H}_2$ symmetry violating operators may give an
appreciable contribution to the amplitude of $K^0-\overline{K}^0$ oscillations and
give rise to new muon decay channels like $\mu\to e^{-}e^{+}e^{-}$. In order to suppress
processes with non--diagonal flavour transitions the Yukawa couplings of the exotic
particles to the quarks and leptons of the first two generations should be smaller
than $10^{-3}-10^{-4}$. Such small $Z^{H}_2$ symmetry violating couplings can be ignored
in the first approximation.

In this approximation and given the previous assumption that only $H_u$, $H_d$
and $S$ acquire non-zero VEVs the charged components of the Inert Higgsinos
$(\tilde{H}^{u+}_2,\,\tilde{H}^{u+}_1,\,\tilde{H}^{d-}_2,\,\tilde{H}^{d-}_1)$
and ordinary chargino states do not mix. The neutral components of the Inert Higgsinos
($\tilde{H}^{d0}_1$, $\tilde{H}^{d0}_2$, $\tilde{H}^{u0}_1$, $\tilde{H}^{u0}_2$)
and Inert singlinos ($\tilde{S}_1$, $\tilde{S}_2$) also do not mix  with
the ordinary neutralino states. Moreover if $Z^{H}_2$ symmetry was exact then both the lightest state in the
ordinary neutralino sector and the lightest Inert neutralino would be absolutely stable. Therefore,
although $Z^{H}_2$ symmetry violating couplings are expected to be rather small, we shall
assume that they are large enough to allow either the lightest neutralino state or the lightest
Inert neutralino to decay within a reasonable time,
the lighter of the two being the stable LSP.

In the field basis
$(\tilde{H}^{d0}_2,\,\tilde{H}^{u0}_2,\,\tilde{S}_2,\,\tilde{H}^{d0}_1,\,\tilde{H}^{u0}_1,\,\tilde{S}_1)$
the mass matrix of the Inert neutralino sector takes a form
\begin{equation}
M_{IN}=
\left(
\begin{array}{cc}
A_{22}  & A_{21} \\[2mm]
A_{12}  & A_{11}
\end{array}
\right)\,,
\label{icn1}
\end{equation}
where $A_{\alpha\beta}$ are $3\times 3$ sub-matrices given by \cite{Hall:2009aj}:
\begin{equation}
A_{\alpha\beta}=-\dfrac{1}{\sqrt{2}}
\left(
\begin{array}{ccc}
0                                           & \lambda_{\alpha\beta} s           & \tilde{f}_{\beta\alpha} v \sin{\beta} \\[2mm]
\lambda_{\beta\alpha} s                     & 0                                 & f_{\beta\alpha} v \cos{\beta} \\[2mm]
\tilde{f}_{\alpha\beta} v \sin{\beta}       & f_{\alpha\beta} v \cos{\beta}     & 0
\end{array}
\right)\,,
\label{icn2}
\end{equation}
so that $A_{\alpha\beta}=A^{T}_{\beta\alpha}$. In the basis of Inert chargino interaction states
$(\tilde{H}^{u+}_2,\,\tilde{H}^{u+}_1,\,\tilde{H}^{d-}_2,\,\tilde{H}^{d-}_1)$
the corresponding mass matrix can be written as
\begin{equation}
M_{IC}=
\left(
\begin{array}{cc}
0  & C^{T} \\[2mm]
C  & 0
\end{array}
\right)\,,\qquad C_{\alpha\beta}=\dfrac{1}{\sqrt{2}}\lambda_{\alpha\beta}\, s\,.
\label{icn3}
\end{equation}
where $C_{\alpha\beta}$ are $2\times 2$ sub-matrices. From Eqs.~(\ref{icn1})--(\ref{icn3}) one can see that
in the exact $Z_2^H$ symmetry limit the spectrum of the Inert neutralinos and charginos in the E$_6$SSM can
be parametrised in terms of
\begin{equation}
\lambda_{\alpha\beta}\,,\qquad f_{\alpha\beta}\,,\qquad \tilde{f}_{\alpha\beta}\,,\qquad \tan\beta\,,\qquad s\,.
\label{icn4}
\end{equation}
In other words the masses and couplings of the Inert neutralinos are determined by 12 Yukawa
couplings, which can be complex, $\tan\beta$ and $s$. Four of the Yukawa couplings mentioned above,
i.e. $\lambda_{\alpha\beta}$, as well as the VEV of the SM singlet field $s$ set the masses and couplings
of the Inert chargino states. Six off--diagonal Yukawa couplings define the mixing between the two families of
the Inert Higgsinos and singlinos.

In the following analysis we shall choose the VEV of the SM singlet field to be large enough
($s\gtrsim 2400\,\mbox{GeV}$) so that the experimental constraints on $Z'$ boson mass
($M_{Z'}\gtrsim 865\,\mbox{GeV}$) and $Z-Z'$ mixing are satisfied. In order to avoid the LEP lower
limit on the masses of Inert charginos the Yukawa couplings $\lambda_{\alpha\beta}$ are chosen
so that all Inert chargino states are heavier than $100\,\mbox{GeV}$. In addition, we also require
the validity of perturbation theory up to the GUT scale and that constrains the allowed range of
all Yukawa couplings.

The theoretical and experimental restrictions specified above set very strong limits on the masses
and couplings of the lightest Inert neutralinos. In particular, our numerical analysis indicates that
the lightest and second lightest Inert neutralinos are always light. They typically have masses
below $60-65\,\mbox{GeV}$. These neutralinos are predominantly Inert singlinos.
From our numerical analysis it follows that the lightest and second lightest Inert neutralinos
might have rather small couplings to the $Z$--boson so that any possible signal which these neutralinos
could give rise to at LEP would be extremely suppressed. As a consequence such Inert neutralinos
would remain undetected. At the same time four other Inert neutralinos, which are approximately linear
superpositions of neutral components of Inert Higgsinos, are normally heavier than $100\,\mbox{GeV}$.

\subsection{The diagonal Inert Yukawa approximation}
\label{diagonalInertYukawas}

In order to clarify the results of our numerical analysis, it is useful to consider a few simple
cases that give some analytical understanding of our calculations. In the simplest
case when all off--diagonal Yukawa couplings vanish, considered in  \cite{Hall:2009aj},
$$
\lambda_{\alpha\beta}=\lambda_{\alpha}\,\delta_{\alpha\beta},\qquad
f_{\alpha\beta}=f_{\alpha}\,\delta_{\alpha\beta},\qquad
\tilde{f}_{\alpha\beta}=\tilde{f}_{\alpha}\,\delta_{\alpha\beta}\,,
$$
the mass matrix of Inert neutralinos reduces to the block diagonal form while the masses of
the Inert charginos are given by
\begin{equation}
m_{\chi^{\pm}_{\alpha}}=\dfrac{\lambda_{\alpha}}{\sqrt{2}}\,s\,.
\label{icn5}
\end{equation}

In the limit where $f_{\alpha}=\tilde{f}_{\alpha}$ one can easily prove using the method
proposed in \cite{Hesselbach:2007te} that there are theoretical upper bounds on the masses
of the lightest and second lightest Inert neutralino states. The corresponding theoretical
restrictions are
\begin{equation}
\begin{array}{l}
|m_{\chi^0_{\alpha}}|^2\lesssim \mu_{\alpha}^2=\ds\frac{1}{2}\biggl[|m_{\chi^{\pm}_{\alpha}}|^2+\ds\frac{f_{\alpha}^2 v^2}{2}
\biggl(1+\sin^22\beta\biggr)-\\[3mm]
\qquad\qquad\qquad\qquad\qquad\qquad\sqrt{\biggl(|m_{\chi^{\pm}_{\alpha}}|^2+\ds\frac{f_{\alpha}^2 v^2}{2}(1+\sin^22\beta)\biggr)^2-
f_{\alpha}^4 v^4 \sin^22\beta}\biggr]\,.
\end{array}
\label{icn6}
\end{equation}
The value of $\mu_{\alpha}$ decreases with increasing $|m_{\chi^{\pm}_{\alpha}}|$ and $\tan\beta$, hence reaching
its maximum value of ${f_\alpha \over\sqrt{2}} v$ for $m_{\chi^{\pm}_{\alpha}}\to 0$ and $\tan\beta\to 1$. At large values of $|m_{\chi^{\pm}_{\alpha}}|$
and $\tan\beta$, Eq.~(\ref{icn6}) simplifies resulting in
\begin{equation}
|m_{\chi^0_{\alpha}}|^2\lesssim \ds\frac{f_{\alpha}^4 v^4 \sin^22\beta}{4\biggl(|m_{\chi^{\pm}_{\alpha}}|^2+
\ds\frac{f_{\alpha}^2 v^2}{2}(1+\sin^22\beta)\biggr)}\,.
\label{icn7}
\end{equation}
Eqs.~(\ref{icn6})-(\ref{icn7}) demonstrate that the upper bound on the mass of the
lightest Inert neutralino also depends on the values of Yukawa couplings $f_{\alpha}$ and $\tilde{f}_{\alpha}$.
For relatively small values of $\tan\beta$, the theoretical restrictions on $f_{\alpha}$ and $\tilde{f}_{\alpha}$, due to the requirement that the
perturbation theory is valid up to the GUT scale, become weaker with increasing $\tan\beta$. However,
at large values of $\tan\beta$ the upper bounds on $|m_{\chi^0_{\alpha}}|$ become rather small according to
Eqs.~(\ref{icn6})-(\ref{icn7}). When $\tan\beta$ tends to unity, $\mu_{\alpha}^2$ also decreases because the
constraints on $f_{\alpha}$ and $\tilde{f}_{\alpha}$ become more and more stringent. The theoretical restrictions on
$|m_{\chi^0_{\alpha}}|$ achieve their maximal value around $\tan\beta\simeq 1.5$. For this value of $\tan\beta$
the requirement of the validity of perturbation theory up to the GUT scale implies that
$f_{1}=\tilde{f}_{1}=f_{2}=\tilde{f}_{2}$ are less than $0.6$. As a consequence the lightest Inert neutralinos are
lighter than $60-65\,\mbox{GeV}$ for $|m_{\chi^{\pm}_{\alpha}}|> 100\,\mbox{GeV}$.

The Inert neutralino mass matrix (\ref{icn1})-(\ref{icn2}) can be diagonalized using the neutralino mixing
matrix defined by
\begin{equation}
N_i^a M^{ab} N_j^b = m_i \delta_{ij}, \qquad\mbox{ no sum on } i.
\label{icn8}
\end{equation}
In the limit where off--diagonal Yukawa couplings vanish and $\lambda_{\alpha} s\gg f_{\alpha} v,\, \tilde{f}_{\alpha} v$
the eigenvalues of the Inert neutralino mass matrix can be easily calculated (see \cite{Hall:2009aj}).
The masses of the four heaviest Inert neutralinos are set by the masses of Inert chargino states
\begin{equation}
m_{\chi^0_{3,4,5,6}}\simeq \pm m_{\chi^{\pm}_{\alpha}}-
\dfrac{\tilde{f}_{\alpha} f_{\alpha} v^2 \sin 2\beta}{4 m_{\chi^{\pm}_{\alpha}}}\,.
\label{icn9}
\end{equation}
The masses of the two lightest Inert neutralinos are determined by the values of the Yukawa couplings
$\tilde{f}_{\alpha}$ and $f_{\alpha}$
\begin{equation}
m_{\chi^0_{\alpha}}\simeq \dfrac{\tilde{f}_{\alpha} f_{\alpha} v^2 \sin 2\beta}{2 m_{\chi^{\pm}_{\alpha}}}\,.
\label{icn10}
\end{equation}
These are naturally small and hence good candidates for being the LSP and NLSP since
$m_{\chi^{\pm}_{\alpha}}\sim s$ from Eq.~(\ref{icn5}) and hence $m_{\chi^0_{\alpha}}\sim v^2/s$
as observed in \cite{Hall:2009aj}.

Again one can see that the masses of the lightest Inert neutralino states decrease with increasing
$\tan\beta$ and chargino masses. In this approximation the lightest Inert neutralinos are
made up of the following superposition of interaction states
\begin{equation}
\tilde{\chi}_{\alpha}^0 = N_{\alpha}^1 \tilde{H}^{d0}_2 + N_{\alpha}^2 \tilde{H}^{u0}_2 + N_{\alpha}^3 \tilde{S}_2 +
N_{\alpha}^4 \tilde{H}^{d0}_1 + N_{\alpha}^5 \tilde{H}^{u0}_1 + N_{\alpha}^6 \tilde{S}_1\,,
\label{icn11}
\end{equation}
where
\begin{equation}
\begin{array}{l}
N_{1}^1=N_{1}^2=N_{1}^3=0\,,\qquad N_{1}^4 \simeq -\dfrac{f_1 v \cos\beta}{\lambda_1 s}\,,\qquad
N_{1}^5 \simeq -\dfrac{\tilde{f}_1 v \sin\beta}{\lambda_1 s}\,,\\[1mm]
N_{1}^6 \simeq 1 - \dfrac{1}{2} \left(\dfrac{v}{\lambda_1 s}\right)^2
\left[f_1^2\cos^2\beta + \tilde{f}_1^2\sin^2\beta\right]\,;\\[1mm]
N_{2}^1 \simeq -\dfrac{f_2 v \cos\beta}{\lambda_2 s}\,,\qquad
N_{2}^2 \simeq -\dfrac{\tilde{f}_2 v \sin\beta}{\lambda_2 s}\,,\qquad N_{2}^4=N_{2}^5=N_{2}^6=0\,,\\[1mm]
N_{2}^3 \simeq 1 - \dfrac{1}{2} \left(\dfrac{v}{\lambda_2 s}\right)^2
\left[f_2^2\cos^2\beta + \tilde{f}_2^2\sin^2\beta\right]\,.
\end{array}
\label{icn12}
\end{equation}
From Eq.~(\ref{icn12}) it becomes clear that the lightest and second lightest Inert neutralinos are
mostly Inert singlinos.

Using the above lightest and second lightest Inert neutralino compositions
it is straightforward to derive the couplings of these states to the $Z$-boson.
In general the part of the Lagrangian that describes the interactions of $Z$ with $\chi^0_1$ and $\chi^0_2$,
can be presented in the following form:
\begin{equation}
\begin{array}{c}
\mathcal{L}_{Z\chi\chi}=\sum_{\alpha,\beta}\dfrac{M_Z}{2 v}Z_{\mu}
\biggl(\chi^{0T}_{\alpha}\gamma_{\mu}\gamma_{5}\chi^0_{\beta}\biggr) R_{Z\alpha\beta}\,,\\[1mm]
R_{Z\alpha\beta}=N_{\alpha}^1 N_{\beta}^1 - N_{\alpha}^2 N_{\beta}^2 + N_{\alpha}^4 N_{\beta}^4 -
N_{\alpha}^5 N_{\beta}^5\,.
\end{array}
\label{icn13}
\end{equation}
In the case where off--diagonal Yukawa couplings go to zero while
$\lambda_{\alpha} s\gg f_{\alpha} v,\, \tilde{f}_{\alpha} v$
the relative couplings of the lightest and second lightest Inert
neutralino states to the $Z$-boson are given by
\begin{equation}
R_{Z\alpha\beta}=R_{Z\alpha\alpha}\,\delta_{\alpha\beta}\,,\qquad
R_{Z\alpha\alpha}=\dfrac{v^2}{2 m_{\chi^{\pm}_{\alpha}}^2}
\biggl(f_{\alpha}^2\cos^2\beta-\tilde{f}_{\alpha}^2\sin^2\beta\biggr)\,.
\label{icn14}
\end{equation}
Eq.~(\ref{icn14}) demonstrates that the couplings of $\chi^0_1$ and $\chi^0_2$ to the $Z$-boson can be very strongly
suppressed or even tend to zero. This happens when $|f_{\alpha}|\cos\beta=|\tilde{f}_{\alpha}|\sin\beta$,
which is when $\chi^0_{\alpha}$ contains a completely symmetric combination of $\tilde{H}^{d0}_{\alpha}$ and
$\tilde{H}^{u0}_{\alpha}$. Eq.~(\ref{icn14}) also indicates that the couplings of $\chi^0_1$ and $\chi^0_2$ to Z
are always small when Inert charginos are rather heavy or $\tilde{f}_{\alpha}$ and $f_{\alpha}$ are small
(i.e. $m_{\chi^0_{\alpha}}\to 0$).

\subsection{$\Delta_{27}$ and pseudo-Dirac lightest neutralino states}
In order to provide an explanation of the origin of the aproximate
$Z_2^H$ symmetry that singles out the third family of Higgs doublets and singlets,
and to account for tri-bimaximal mixing and other features of the quark and lepton
spectrum, a $\Delta_{27}$ family symmetry has been applied to the E$_6$SSM
\cite{Howl:2009ds} \footnote{The corresponding mass terms come from
the product {\bf $(3\times 3\times 3_f)(3\times \bar{3}'_f)$},
where {\bf $3$, $3_f$} and {\bf $\bar{3}'_f$} are triplet representations of $\Delta_{27}$.
The $27_i$ multiplets that contain quarks and leptons form $3$ representations of the $\Delta_{27}$ group.
{\bf $3_f$} and {\bf $\bar{3}'_f$} contain flavon fields that break $\Delta_{27}$. In the considered
model the non-zero mass of the lightest Inert neutralino state is induced by the symmetric
invariant that appears in the {\bf $(3\times 3\times 3)$} decomposition of the $\Delta_{27}$ triplet
representation (i.e. $123+231+312+213+321+132$) \cite{Ma:2006ip}.}.
The addition of a $\Delta_{27}$ family symmetry implies
an Inert neutralino mass matrix with $A_{11}\approx A_{22} \approx 0$, where $A_{\alpha \beta}$ are defined in
Eq.~(\ref{icn1}), leading to approximately degenerate lightest neutralino states with a pseudo-Dirac
structure.

When all flavour diagonal Yukawa couplings $\lambda_{\alpha\alpha}$, $f_{\alpha\alpha}$
and $\tilde{f}_{\alpha\alpha}$ exactly vanish, i.e. $A_{11}=A_{22}=0$,
all Inert Higgsinos and singlinos
form Dirac states. In this limit the Lagrangian of the E$_6$SSM is invariant under
a $U(1)$ global symmetry. The fermion components of the Inert Higgs superfields transform
under this symmetry as follows:
\begin{equation}
\begin{array}{lll}
\tilde{S}_1\to e^{i\alpha} \tilde{S}_1\,,\qquad &
\tilde{H}^u_1\to e^{i\alpha} \tilde{H}^u_1\,,\qquad &
\tilde{H}^d_1\to e^{i\alpha} \tilde{H}^d_1\,,\\
\tilde{S}_2\to e^{-i\alpha} \tilde{S}_2\,,\qquad &
\tilde{H}^u_2\to e^{-i\alpha} \tilde{H}^u_2\,,\qquad &
\tilde{H}^d_2\to e^{-i\alpha} \tilde{H}^d_2\,.
\end{array}
\label{icn19}
\end{equation}
In the above limiting case
the lightest Inert neutralino is a Dirac state formed predominantly by $\tilde{S}_1$ and $\tilde{S}_2$.
In this case the LSP and its antiparticle have opposite charges with respect to the extra global
$U(1)$ and this might lead to so-called asymmetric dark matter (ADM) \cite{adm}--\cite{adm1}.
In the framework of the ADM scenario there can be an asymmetry between the density of dark matter particles
and their antiparticles in the early universe similar to that for ordinary baryons. This may have a considerable
effect on the relic density calculations \cite{Griest:1986yu}. In particular, if an asymmetry exists between
the number density of dark matter particles and their antiparticles in the early universe, then one can get an
appreciable dark matter density even if the dark matter particle--antiparticle annihilation cross section is very
large like in the case of baryons. Moreover if most of the dark matter antiparticles are eliminated by annihilation
with their particles then such an ADM scenario does not have the usual indirect signatures associated with the
presence of dark matter (e.g. there is no high energy neutrino signal from annihilations in the Sun etc.).
At the same time, a relatively high concentration of dark matter particles can build up in the Sun
altering heat transport in the solar interior and affecting the low energy neutrino fluxes \cite{adm1}.

In practice the $\Delta_{27}$ scenario tells us that we are somewhat away from the above limiting case,
with a broken global $U(1)$ symmetry
leading to almost degenerate pseudo-Dirac lightest neutralinos, where the relic density
of the LSP can be calculated by standard methods.
It will turn out that the LSP cannot be too light (must be of order $M_Z/2$) in order not
to have too high a cosmological relic density. At the same time we will see that the two lightest neutralinos
cannot be too heavy in order for perturbation theory to be valid up to the GUT scale.
In practice this means that in realistic scenarios the two lightest Inert neutralino states
are rather close in mass. The $\Delta_{27}$ scenario provides a natural explanation
of this successful neutralino mass pattern. It is worth noting that the results from the previous
section can be reinterpreted in terms of this scenario. Specifically in the case where
$A_{11} = A_{22} = 0$ and $A_{21} = A_{12}$ a block diagonalisation of the Inert neutralino mass matrix
(\ref{icn1}) yields $A_{22} \rightarrow A_{22}' = -A_{21}$ and $A_{11} \rightarrow A_{11}' = A_{21}$
(with $A_{21} = A_{12} \rightarrow A_{21}' = A_{12}' = 0$).
This only corresponds to a redefinition of the generations 1 and 2 and does not mix
fields of different hypercharge. This provides the following dictionary between these two scenarios:
$\lambda_{11}' = -\lambda_{22}' = \lambda_{21}$; $f_{11}' = -f_{22}' = f_{21}$;
$\tilde{f}_{11}' = -\tilde{f}_{22}' = \tilde{f}_{21}$.
Rewriting the Inert neutralino mass matrix in this block diagonal form also makes
it clear that the $R_{Z12}$ coupling vanishes in this limit, as it did
in the subsection (\ref{diagonalInertYukawas}).



\subsection{Scenario with one light family of Inert Higgsinos}

Another limit that it is worth considering corresponds to the case where one pair of Inert
Higgs doublets decouples from the rest of the spectrum. This occurs when either the corresponding
states are extremely heavy ($\gtrsim 1\,\mbox{TeV}$) or they have rather small couplings to
other Inert Higgs fields. When $\tilde{H}^{d0}_2$ and $\tilde{H}^{u0}_2$ decouple,
the Inert neutralino mass matrix (\ref{icn1}) reduces to a $4\times 4$ matrix. If
$\lambda_{11} s\gg f_{\alpha 1} v,\, \tilde{f}_{\alpha 1} v$, the Inert Higgs states
associated with $\tilde{H}^{d0}_1$ and $\tilde{H}^{u0}_1$ can be integrated out. Then
the resulting $2\times 2$ mass matrix can be written as follows
\begin{equation}
M_{IS}=
\dfrac{v^2\sin 2\beta}{4 m_{\chi^{\pm}_1}}
\left(
\begin{array}{cc}
2\tilde{f}_{11} f_{11}                         & \tilde{f}_{11} f_{21} + f_{11} \tilde{f}_{21}\\[2mm]
\tilde{f}_{11} f_{21} + f_{11} \tilde{f}_{21}  & 2\tilde{f}_{21} f_{21}
\end{array}
\right)\,.
\label{icn15}
\end{equation}
The masses of the lightest and second lightest Inert neutralinos, which are predominantly
superpositions of the Inert singlinos $\tilde{S}_1$ and $\tilde{S}_2$, are given by
\begin{equation}
m_{\chi^0_1,\,\chi^0_2}=
\dfrac{v^2\sin 2\beta}{4 m_{\chi^{\pm}_1}}\biggl[
\tilde{f}_{11} f_{11}+\tilde{f}_{21} f_{21}\pm\sqrt{(f_{11}^2+f_{21}^2)(\tilde{f}_{11}^2+\tilde{f}_{21}^2)}
\biggr]\,.
\label{icn16}
\end{equation}
From Eq.~(\ref{icn16}) it is easy to see that the substantial masses of
the lightest and second lightest Inert neutralinos can be induced even if only one
family of the Inert Higgsinos couples to $S_1$ and $S_2$.

Using Eq.~(\ref{icn13}) one can also calculate the couplings of $\chi^0_1$ and $\chi^0_2$
to the Z-boson
\begin{equation}
\begin{array}{rcl}
R_{Z11}&=&\dfrac{v^2}{2 m^2_{\chi^{\pm}_1}}\biggl[(f_{11}\cos\theta+f_{21}\sin\theta)^2\cos^2\beta-
(\tilde{f}_{11}\cos\theta+\tilde{f}_{21}\sin\theta)^2\sin^2\beta\biggr]\,,\\[1mm]
R_{Z22}&=&\dfrac{v^2}{2 m^2_{\chi^{\pm}_1}}\biggl[(f_{21}\cos\theta-f_{11}\sin\theta)^2\cos^2\beta-
(\tilde{f}_{21}\cos\theta-\tilde{f}_{11}\sin\theta)^2\sin^2\beta\biggr]\,,\\[1mm]
R_{Z12}&=&R_{Z21}=\dfrac{v^2}{2 m^2_{\chi^{\pm}_1}}\biggl[\left(\dfrac{1}{2}(f_{21}^2-f^2_{11})\sin 2\theta+
f_{11} f_{21}\cos 2\theta\right)\cos^2\beta\\[1mm]
&&\qquad\qquad\qquad\qquad-\left(\dfrac{1}{2}(\tilde{f}_{21}^2-\tilde{f}^2_{11})\sin 2\theta+
\tilde{f}_{11} \tilde{f}_{21}\cos 2\theta\right)\sin^2\beta\biggr]\,,
\end{array}
\label{icn17}
\end{equation}
where
$\tan 2\theta=(\tilde{f}_{11} f_{21}+\tilde{f}_{21} f_{11})/(\tilde{f}_{21} f_{21}-\tilde{f}_{11} f_{11})$.
Again from Eqs.~(\ref{icn17}) it follows that $R_{Z11}$, $R_{Z22}$ and $R_{Z12}$
are typically small since $m_{\chi^{\pm}_{\alpha}}\sim s$ from Eq.~(\ref{icn5}) and hence
they are proportional to $v^2/s^2$. However this assumes the lightest Inert chargino is rather heavy.
Alternatively the couplings may be small due to a cancellation between different contributions in
Eqs.~(\ref{icn17}), and/or the f-couplings being small (i.e. $m_{\chi^0_1,\,\chi^0_2}\to 0$).

The simple hierarchical structure of the spectrum of the Inert neutralinos considered above
allows us to highlight an interesting scenario which does not normally appear in the simplest
SUSY extensions of the SM such as the MSSM and NMSSM. When $\tilde{f}_{11}=f_{21}=0$ the diagonal
entries of the mass matrix (\ref{icn15}) vanish leading to the formation of a Dirac lightest
Inert neutralino state. In this case the Lagrangian of the model is invariant under extra
$U(1)$ global symmetry transformations
\footnote{Similar results can be obtained for $f_{11}=\tilde{f}_{21}=0$}
$\tilde{S}_1\to e^{i\alpha} \tilde{S}_1$\,,
$\tilde{H}^u_1\to e^{-i\alpha} \tilde{H}^u_1$\,,
$\tilde{H}^d_1\to e^{i\alpha} \tilde{H}^d_1$\,,
$\tilde{S}_2\to e^{-i\alpha} \tilde{S}_2$\,.
In fact if the E$_6$SSM possess such an exact $U(1)$ global symmetry,
then the spectrum of the Inert neutralinos contains a set of Dirac states only.

\section{Higgs masses and couplings}

The presence of light Inert neutralinos in the particle spectrum of the E$_6$SSM makes possible the
decays of the Higgs bosons into these exotic final states. In this and the next section we argue that
such decays may result in the modification of the SM-like Higgs signal at current and future colliders.
Since our main concern in this paper is the decays of the SM-like lightest Higgs boson, we shall
ignore the effects of the Inert Higgs scalars and pseudoscalars which do not mix appreciably with the
scalar sector responsible for EWSB.
We also assume that all the Inert bosons are heavier than the SM-like Higgs boson.

The sector responsible for the EWSB in the E$_6$SSM includes two Higgs doublets $H_u$ and $H_d$ as
well as the SM singlet field $S$. The Higgs effective potential can be written in the following form:
\be
\ba{rcl}
V&=&V_F+V_D+V_{soft}+\Delta V\, ,\\[2mm]
V_F&=&\lambda^2|S|^2(|H_d|^2+|H_u|^2)+\lambda^2|(H_d H_u)|^2\,,\\[2mm]
V_D&=&\ds\frac{g_2^2}{8}\left(H_d^\dagger \sigma_a H_d+H_u^\dagger \sigma_a
H_u\right)^2+\frac{{g'}^2}{8}\left(|H_d|^2-|H_u|^2\right)^2+\\[2mm]
&&+\ds\frac{g^{'2}_1}{2}\left(\tilde{Q}_1|H_d|^2+\tilde{Q}_2|H_u|^2+\tilde{Q}_S|S|^2\right)^2\,,\\[2mm]
V_{soft}&=&m_{S}^2|S|^2+m_1^2|H_d|^2+m_2^2|H_u|^2+\biggl[\lambda A_{\lambda}S(H_u H_d)+h.c.\biggr]\,,
\ea
\label{higgs1}
\ee
where $g_2$, $g'=\sqrt{3/5} g_1$ and $g^{'}_1$ are the low energy $SU(2)_W$, $U(1)_Y$ and $U(1)_{N}$
gauge couplings while $\tilde{Q}_1$, $\tilde{Q}_2$ and $\tilde{Q}_S$ are the effective $U(1)_{N}$ charges
of $H_d$, $H_u$ and $S$. The term $\Delta V$ represents the contribution from loop corrections to the
Higgs effective potential. Here $H_d^T=(H_d^0,\,H_d^{-})$, $H_u^T=(H_u^{+},\,H_u^{0})$ and
$(H_d H_u)=H_u^{+} H_d^{-} - H_u^{0} H_d^{0}$.

Initially the EWSB sector involves ten degrees of freedom. However four of them are massless Goldstone
modes which are swallowed by the $W^{\pm}$, $Z$ and $Z'$ gauge bosons that gain non-zero masses.
In the limit where $s\gg v$ the masses of the $W^{\pm}$, $Z$ and $Z'$ gauge bosons are given by
$$
M_W=\ds\frac{g_2}{2}v\,,\qquad M_Z\simeq\ds\frac{\bar{g}}{2}v\,,\qquad M_{Z'}\simeq g'_1 \tilde{Q}_S\, s\,,
$$
where $\bar{g}=\sqrt{g_2^2+g'^2}$. When CP--invariance is preserved the other degrees of freedom form
two charged, one CP--odd and three CP-even Higgs states. The masses of the charged and CP-odd Higgs bosons are
\be
m^2_{H^{\pm}}=\ds\frac{\sqrt{2}\lambda A_{\lambda}}{\sin 2\beta}s-\frac{\lambda^2}{2}v^2+M_W^2+\Delta_{\pm}\,,\qquad\quad
m^2_{A}\simeq \ds\frac{\sqrt{2}\lambda A_{\lambda}}{\sin 2\beta}s+\Delta_A\,,
\label{higgs2}
\ee
where $\Delta_{\pm}$ and $\Delta_A$ are the loop corrections.

The CP--even Higgs sector involves $Re\,H_d^0$, $Re\,H_u^0$ and $Re\,S$. In the field space basis $(h,\,H,\,N)$,
rotated by an angle $\beta$ with respect to the initial one,
\be
\ba{c}
Re\,H_d^0=(h \cos\beta- H\sin\beta+v_1)/\sqrt{2}\,, \\[2mm]
Re\,H_u^0=(h \sin\beta+ H\cos\beta+v_2)/\sqrt{2}\,, \\[2mm]
Re\,S=(s+N)/\sqrt{2}\,,
\ea
\label{higgs3}
\ee
the mass matrix of the CP--even Higgs sector takes the form \cite{Nevzorov:2001um}:
\be
M^2=
\left(
\ba{ccc}
\ds\frac{\partial^2 V}{\partial v^2}&
\ds\frac{1}{v}\frac{\partial^2 V}{\partial v \partial\beta}&
\ds\frac{\partial^2 V}{\partial v \partial s}\\[0.3cm]
\ds\frac{1}{v}\frac{\partial^2 V}{\partial v \partial\beta}&
\ds\frac{1}{v^2}\frac{\partial^2 V}{\partial^2\beta}&
\ds\frac{1}{v}\frac{\partial^2 V}{\partial s \partial\beta}\\[0.3cm]
\ds\frac{\partial^2 V}{\partial v \partial s}&
\ds\frac{1}{v}\frac{\partial^2 V}{\partial s \partial\beta}&
\ds\frac{\partial^2 V}{\partial^2 s}
\ea
\right)=\left(
\ba{ccc}
M_{11}^2 & M_{12}^2 & M_{13}^2\\
M_{21}^2 & M_{22}^2 & M_{23}^2\\
M_{31}^2 & M_{32}^2 & M_{33}^2
\ea
\right)\,,
\label{higgs4}
\ee
where
$$
\ba{rcl}
M_{11}^2&=&\ds\frac{\lambda^2}{2}v^2\sin^22\beta+\ds\frac{\bar{g}^2}{4}v^2\cos^22\beta+g^{'2}_1 v^2(\tilde{Q}_1\cos^2\beta+
\tilde{Q}_2\sin^2\beta)^2+\Delta_{11}\,,\\[2mm]
M_{12}^2&=&M_{21}^2=\ds\left(\frac{\lambda^2}{4}-\ds\frac{\bar{g}^2}{8}\right)v^2
\sin 4\beta+\ds\frac{g^{'2}_1}{2}v^2(\tilde{Q}_2-\tilde{Q}_1)\times\\[2mm]
&&\times(\tilde{Q}_1\cos^2\beta+\tilde{Q}_2\sin^2\beta)\sin 2\beta+\Delta_{12}\, ,\\
M_{22}^2&=&\ds\frac{\sqrt{2}\lambda A_{\lambda}}{\sin 2\beta}s+\left(\frac{\bar{g}^2}{4}-\ds\frac{\lambda^2}{2}\right)v^2
\sin^2 2\beta+\ds\frac{g^{'2}_1}{4}(\tilde{Q}_2-\tilde{Q}_1)^2 v^2 \sin^22\beta+\Delta_{22}\,,
\ea
$$
\be
\ba{rcl}
M_{23}^2&=&M_{32}^2=-\ds\frac{\lambda A_{\lambda}}{\sqrt{2}}v\cos 2\beta+\ds\frac{g^{'2}_1}{2}(\tilde{Q}_2-\tilde{Q}_1)\tilde{Q}_S
v s\sin 2\beta+\Delta_{23}\,,\\
M_{13}^2&=&M_{31}^2=-\ds\frac{\lambda A_{\lambda}}{\sqrt{2}}v\sin 2\beta+\lambda^2 v s+g^{'2}_1(\tilde{Q}_1\cos^2\beta+
\tilde{Q}_2\sin^2\beta)\tilde{Q}_S v s+\Delta_{13}\,,\\
M_{33}^2&=&\ds\frac{\lambda A_{\lambda}}{2\sqrt{2}s}v^2\sin 2\beta+M_{Z'}^2+\Delta_{33}\,.
\ea
\label{higgs5}
\ee
In Eq.~(\ref{higgs5}) the $\Delta_{ij}$ represent the contributions from loop corrections which in the leading one--loop
approximation are rather similar to the ones calculated in the NMSSM\footnote{Note that the explicit expressions
for $\Delta_{ij}$, $\Delta_{\pm}$ and $\Delta_A$ presented in the first paper in \cite{Nevzorov:2001um} contain
a typo. In the corresponding formulae $\mu$ is neither a parameter of the MSSM Lagrangian nor an effective $\mu$--term
in the NMSSM. It has to be associated with the renormalisation scale.}. Since the minimal eigenvalue of the mass
matrix (\ref{higgs4})--(\ref{higgs5}) is always less than its smallest diagonal element, at least one Higgs
scalar in the CP--even sector (approximately $h$) remains always light, i.e. $m^2_{h_1}\lesssim M_{11}^2$.
In the leading two--loop approximation the mass of the lightest Higgs boson in the E$_6$SSM does not exceed
$150-155\,\mbox{GeV}$. When the SUSY breaking scale $M_S$ and the VEV $s$ of the singlet field are considerably larger
than the EW scale, the mass matrix (\ref{higgs4})--(\ref{higgs5}) has a hierarchical structure and can be
diagonalised using the perturbation theory \cite{Nevzorov:2001um}-\cite{Nevzorov:2004ge}. In this case
the masses of the heaviest Higgs bosons are closely approximated by the diagonal entries $M_{22}^2$ and
$M_{33}^2$ \cite{King:2005jy}. As a result the mass of one CP--even Higgs boson (approximately given by $H$)
is governed by $m_A$ while the mass of another one (predominantly the $N$ singlet field) is set by $M_{Z'}$.
When $\lambda\gtrsim g'_1$, vacuum stability requires $m_A$ to be considerably larger than $M_{Z'}$ and
the EW scale so that the qualitative pattern of the Higgs spectrum is rather similar to the one which arises
in the PQ symmetric NMSSM \cite{Nevzorov:2004ge}-\cite{Miller:2005qua}. In the considered limit the heaviest
CP--even, CP--odd and charged states are almost degenerate around $m_A$ and lie beyond the $\mbox{TeV}$
range \cite{King:2005jy}.

If all other Higgs states are much heavier than the lightest CP-even Higgs boson then the lightest Higgs state
(approximately given by $h$) manifests itself in the interactions with gauge bosons and fermions as a
SM--like Higgs boson. Since within the E$_6$SSM the mass of this state is predicted to be relatively
low its production cross section at the LHC should be large enough so that it can be observed in the
near future. In this context it is particularly interesting and important to analyse the decay modes of
the lightest CP-even Higgs state. Furthermore we concentrate on the decays of the SM--like Higgs boson into
the lightest and second lightest Inert neutralinos.

The couplings of the Higgs states to the Inert neutralinos originate from the interactions of $H_u$, $H_d$
and $S$ with the Inert Higgs superfields in the superpotential. Using Eqs.~(\ref{higgs3}) one can express
$Re\,H_d^0$, $Re\,H_u^0$ and $Re\,S$ in terms of the components of the CP--even Higgs basis $h$, $H$ and
$N$. At the same time the components of the CP--even Higgs basis are related to the physical CP--even Higgs
eigenstates by virtue of a unitary transformation:
\be
\left(
\begin{array}{c}
h\\ H\\ N
\end{array}
\right)=
U^{\dagger}
\left(
\begin{array}{c}
h_1 \\ h_2\\ h_3
\end{array}
\right)\,.
\label{higgs6}
\ee
Combining all these expressions together one obtains an effective Lagrangian that describes
the interactions of the Inert neutralinos with the CP-even Higgs eigenstates
\be
\ba{l}
\mathcal{L}_{H\chi\chi}=\sum_{i,j,m} (-1)^{\theta_i+\theta_j} X^{h_m}_{ij} \biggl(\psi^{0T}_{i}
(-i\gamma_{5})^{\theta_i+\theta_j}\psi^0_{j}\biggr) h_m\,,\\[4mm]
X^{h_m}_{ij}=-\dfrac{1}{\sqrt{2}}U^{\dagger}_{Nh_{m}}\Lambda_{ij}
-\dfrac{1}{\sqrt{2}}\biggl(U^{\dagger}_{hh_{m}}\cos\beta-U^{\dagger}_{Hh_{m}}\sin\beta\biggr) F_{ij}\\[2mm]
\qquad\qquad\qquad-\dfrac{1}{\sqrt{2}}\biggl(U^{\dagger}_{hh_{m}}\sin\beta+U^{\dagger}_{Hh_{m}}\cos\beta\biggr)\tilde{F}_{ij}\,,\\[4mm]
F_{ij} = f_{11} N^6_i N^5_j + f_{12} N^6_i N^2_j + f_{21} N^3_i N^5_j +f_{22} N^3_i N^2_j\,, \\[2mm]
\tilde{F}_{ij} = \tilde{f}_{11} N^6_i N^4_j + \tilde{f}_{12} N^6_i N^1_j + \tilde{f}_{21} N^3_i N^4_j + \tilde{f}_{22} N^3_i N^1_j\,,\\[2mm]
\Lambda_{ij} = \lambda_{11} N^4_i N^5_j + \lambda_{12} N^4_i N^2_j +\lambda_{21} N^1_i N^5_j + \lambda_{22} N^1_i N^2_j\,,
\ea
\label{higgs7}
\ee
where $i,j=1,2,...6$ and $m=1,2,3$. In Eq.~(\ref{higgs7}) $\psi^0_i=(-i\gamma_5)^{\theta_i}\chi^0_i$ is
the set of Inert neutralino eigenstates with positive eigenvalues, while $\theta_i$ equals 0 (1) if the eigenvalue
corresponding to $\chi^0_i$ is positive (negative). As before, the Inert neutralinos are labeled according to increasing
absolute value of mass, with $\psi^0_1$ being the lightest Inert neutralino and $\psi^0_6$ the heaviest.

The expressions for the couplings of the Higgs scalars to the Inert neutralinos (\ref{higgs7}) become
much more simple in the case of the hierarchical structure of the Higgs spectrum. In this case
$U_{ij}$ is almost an identity matrix. As a consequence, the couplings of the SM-like Higgs boson to
the lightest and second lightest Inert neutralino states are approximately given by
\be
X^{h_1}_{\gamma\sigma}=-\dfrac{1}{\sqrt{2}}\biggl(F_{\gamma\sigma}\cos\beta+\tilde{F}_{\gamma\sigma}\sin\beta\biggr)\,,
\label{higgs8}
\ee
where $\gamma,\sigma=1,2$, labeling the two light, mostly Inert singlino states.
In the limit when off-diagonal Yukawa couplings that determine the interactions
of the inert Higgs fields with $H_u$, $H_d$ and $S$ vanish, as defined in
subsection~(\ref{diagonalInertYukawas}), and Inert neutralino mass matrix has a hierarchical
structure (i.e. $\lambda_{\alpha} s\gg f_{\alpha} v,\, \tilde{f}_{\alpha} v$), one can use the expressions (\ref{icn12})
for $N^{a}_{1,2}$ in order to derive the approximate analytical formulae for $X^{h_1}_{\gamma\sigma}$.
Substituting Eqs.~(\ref{icn12}) into (\ref{higgs8}) one obtains
\be
X^{h_1}_{\gamma\sigma}\simeq\dfrac{|m_{\chi^0_{\sigma}}|}{v}\,\delta_{\gamma\sigma}\,,
\label{higgs9}
\ee
These simple analytical expressions for the couplings of the SM--like Higgs boson to the lightest and
second lightest Inert neutralinos are not as surprising as they may first appear.
When the Higgs spectrum is hierarchical, the VEV of the lightest CP--even state is responsible for
all light fermion masses in the E$_6$SSM. As a result we expect that their couplings to SM--like Higgs can be
written as usual as being proportional to the mass divided by the VEV. We see that this is exactly what is found
in the limit of $|m_{\chi^0_{\sigma}}|$ being small.


\section{Novel Higgs decays and Dark Matter}
\subsection{Higgs decay widths}

The interaction Lagrangian (\ref{higgs7}) gives rise to decays of the lightest Higgs boson into
Inert neutralino pairs with partial widths given by
\be
\ba{rcl}
\Gamma(h_1\to\chi^0_{\alpha}\chi^0_{\beta})&=&\dfrac{\Delta_{\alpha\beta}}{8\pi m_{h_1}}
\biggl(X^{h_1}_{\alpha\beta}+X^{h_1}_{\beta\alpha}\biggr)^2\biggl[
m^2_{h_1}-(|m_{\chi^0_{\alpha}}|+(-1)^{\theta_{\alpha}+\theta_{\beta}}|m_{\chi^0_{\beta}}|)^2\biggr]\\[4mm]
&&\times\sqrt{\biggl(1-\dfrac{|m_{\chi^0_{\alpha}}|^2}{m^2_{h_1}}-\dfrac{|m_{\chi^0_{\beta}}|^2}{m^2_{h_1}}\biggr)^2-
4\dfrac{|m_{\chi^0_{\alpha}}|^2 |m_{\chi^0_{\beta}}|^2}{m^4_{h_1}}}\,,
\ea
\label{higgs10}
\ee
where $\Delta_{\alpha\beta}=\dfrac{1}{2}\,(1)$ for $\alpha=\beta$ ($\alpha\neq\beta$).

The partial widths associated with the exotic decays of the SM-like Higgs boson (\ref{higgs10})
have to be compared with the Higgs decay rates into the SM particles. When the SM-like Higgs state
is relatively light ($m_{h_1}\lesssim 140\,\mbox{GeV}$) it decays predominantly into $b$-quark and
$\tau$--lepton pairs. The partial decay width of the lightest CP--even
Higgs boson into fermion pairs is given by (for recent review see \cite{Djouadi:2005gj})
\be
\Gamma(h_1\to f\bar{f})=N_c\dfrac{g_2^2}{32\pi}\biggl(\dfrac{m_f}{M_W}\biggr)^2 g^2_{h_1ff} m_{h_1}
\biggl(1-\dfrac{4 m_f^2}{m_{h_1}^2}\biggr)^{3/2}\,.
\label{higgs11}
\ee
Eq.~(\ref{higgs11}) can be used for the calculation of the lightest Higgs decay rate into $\tau$--lepton pairs.
In this case the coupling of the lightest CP--even Higgs state to the $\tau$--lepton normalized to the
corresponding SM coupling, i.e. $g_{h_1\tau\tau}$, is given by
\be
g_{h_1\tau\tau}=\dfrac{1}{\cos\beta}\biggl(U^{\dagger}_{hh_1}\cos\beta-U^{\dagger}_{Hh_1}\sin\beta\biggr)\,.
\label{higgs12}
\ee

For a final state that involves $b$--quarks one has to include the QCD corrections. In particular,
the fermion mass in Eq.~(\ref{higgs11}) should be associated with the running $b$--quark mass $\overline{m}_b(\mu)$.
The bulk of the QCD corrections are absorbed by using the running $b$--quark mass defined at the appropriate
renormalisation scale, i.e. at the scale of the lightest Higgs boson mass ($\mu=m_{h_1}$) in the
considered case. In addition to the corrections which are associated with the running $b$--quark mass
there are other QCD corrections to the Higgs coupling to the $b$--quark that should be taken into account
\cite{Gorishnii:1990zu}. As a consequence, the partial decay width of the lightest CP--even Higgs boson into
$b$--quark pairs can be calculated using Eq.~(\ref{higgs11}) if one sets $N_c=3$ and replaces
\be
\ba{l}
m_f\to \overline{m}_b(m_{h_1})\,,\\[2mm]
g^2_{h_1ff}\to \dfrac{1}{\cos^2\beta}\biggl(U^{\dagger}_{hh_1}\cos\beta-U^{\dagger}_{Hh_1}\sin\beta\biggr)^2
\biggl[1+\Delta_{bb}+\Delta_H\biggr]\,,\\[2mm]
\Delta_{bb}\simeq 5.67\dfrac{\bar{\alpha}_s}{\pi}+(35.94-1.36 N_f)\dfrac{\bar{\alpha}^2_s}{\pi^2}\,,\\[2mm]
\Delta_H\simeq \dfrac{\bar{\alpha}^2_s}{\pi^2}\biggl(1.57-\dfrac{2}{3}\log\dfrac{m_{h_1}^2}{m_t^2}+
\dfrac{1}{9}\log^2\dfrac{\overline{m}_b^2}{m_{h_1}^2}\biggr)\,,
\ea
\label{higgs13}
\ee
where $\bar{\alpha}_s=\alpha_s(m_{h_1}^2)$. Here we neglect radiative corrections that originate from
loop diagrams that contain SUSY and exotic particles \footnote{Radiative corrections that are induced
by SUSY particles can be very important particularly in the case of the bottom quark at high values of
$\tan\beta$ (for a review, see \cite{Heinemeyer:2004ms}).}.

From Eqs.~(\ref{higgs9})--(\ref{higgs11}) one can see that in the E$_6$SSM the branching ratios
of the SM--like Higgs state into the lightest and second lightest Inert
neutralinos depend rather strongly on the masses of these exotic particles. When the lightest Inert
neutralino states are relatively heavy, i.e. $m_{\chi_1,\,\chi_2}\gtrsim \overline{m}_b(m_{h_1})$, the
lightest Higgs boson decays predominantly into $\chi_{\alpha}\chi_{\beta}$ while the branching ratios
for decays into SM particles are suppressed. On the other hand if the lightest Inert
neutralinos have masses which are considerably smaller than the masses of the $b$--quark and $\tau$--lepton
then the branching ratios of the exotic decays of the SM--like Higgs state are small. In the E$_6$SSM
the lightest and second lightest Inert neutralinos are expected to be heavier than a few MeV so that
they would not contribute to the expansion rate prior to nucleosynthesis and thus not modify
Big Bang nucleosynthesis (BBN).

\subsection{Dark matter}

More stringent constraints on the masses of the lightest Inert neutralino can be obtained if we require
that this exotic state accounts for all or some of the observed dark matter relic density
which is measured to be $\Omega_{\mathrm{CDM}}h^2 = 0.1099 \pm 0.0062$ \cite{cdm}. If a theory predicts
a greater relic density of dark matter than this then it is ruled out, assuming standard pre-BBN
cosmology. A theory that predicts less dark matter cannot be ruled out in the same way but then
there would have to be other contributions to the dark matter relic density.

In the limit where all non-SM fields other than the two lightest Inert neutralinos
are heavy ($\gtrsim \mbox{TeV}$) the lightest Inert neutralino state in the E$_6$SSM results in too large
a density of dark matter.  As we noted in Section 3, $\tilde{\chi}^0_1$ is usually composed of
Inert singlino and has a mass (Eq. (16)) which is inversely proportional to the charged Higgsino mass.
Thus in this limit it is typically very light $|m_{\chi^0_{\sigma}}|\ll M_Z$. As a result
the couplings of the lightest Inert neutralino to gauge bosons, the SM-like Higgs state, quarks and leptons
are quite small leading to a relatively small annihilation cross section for
$\tilde{\chi}^0_1\tilde{\chi}^0_1\to \mbox{SM particles}$. Since the dark matter number density is inversely
proportional to the annihilation cross section at the freeze-out temperature (see, for example \cite{Wells:1997ag})
the lightest Inert neutralino state gives rise to a relic density that is typically much larger
than its measured value. Thus in the limit considered the bulk of the E$_6$SSM parameter space that
leads to small masses of $\tilde{\chi}^0_1$ is ruled out.

The situation changes dramatically when the mass of the lightest Inert neutralino increases.
In this case the Higgsino components of $\tilde{\chi}^0_1$ become larger and as a consequence
the couplings of $\tilde{\chi}^0_1$ to the $Z$--boson grow \cite{Hall:2009aj}. A reasonable density
of dark matter can be obtained for $|m_{\chi^0_{\sigma}}|\sim M_Z/2$ when the lightest Inert
neutralino states annihilate mainly through an $s$--channel $Z$--boson, via its Inert Higgsino
doublet components which couple to the $Z$--boson. It is worth noting that if
$\tilde{\chi}^0_1$ was pure Inert Higgsino then the $s$--channel $Z$--boson annihilation would proceed
with the full gauge coupling strength leaving the relic density too low to account for the
observed dark matter. In the E$_6$SSM the LSP is mostly Inert singlino so that its coupling
to the $Z$--boson is typically suppressed, since it only couples through its Inert
Higgsino admixture leading to an increased relic density. In practice, the appropriate value
of $\Omega_{\mathrm{CDM}}h^2$ can be achieved even if the coupling of $\tilde{\chi}^0_1$ to
the $Z$--boson is relatively small. This happens when $\tilde{\chi}^0_1$ annihilation proceeds
through the $Z$--boson resonance, i.e. $2|m_{\chi^0_{\sigma}}|\simeq M_Z$ \cite{Hall:2009aj,Barger:2004bz}.
Thus scenarios which result in a reasonable dark matter density correspond to lightest Inert
neutralino masses that are much larger than $\overline{m}_b(m_{h_1})$, and hence the
SM--like Higgs has very small branching ratios into SM particles.

\section{Benchmarks, constraints and predictions}

\begin{figure}[t]
\begin{center}
\includegraphics[width=150mm]{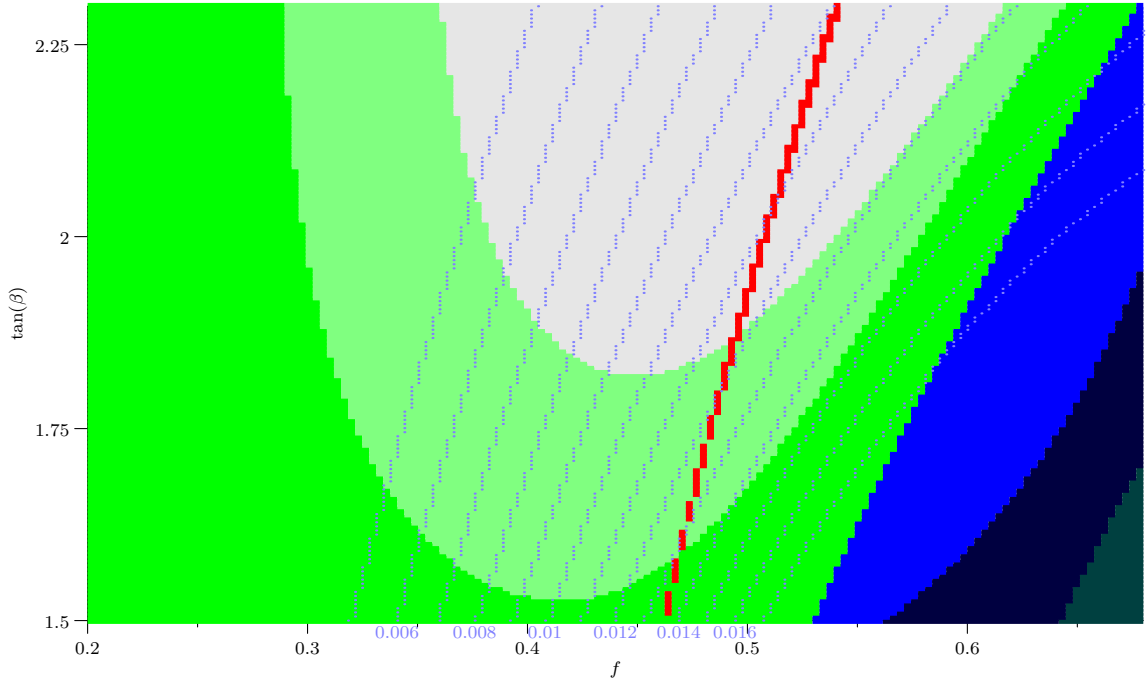}\end{center}
\caption{\it\small Contour plot of $(X^{h_1}_{11})^2$ and relic density
$\Omega_\chi h^2$ regions in the $(f,\tan \beta)$-plane with $s=2400\,\rm{GeV}$,
$f_{\alpha\alpha}=\tilde{f}_{\alpha\alpha}=\lambda_{\alpha\alpha}=0$,
$f_{12}=f$, $\tilde{f}_{12}=f_{12}/a$, $f_{21}=1.02\cdot f_{12}$,
$\tilde{f}_{21}=0.98\cdot \tilde{f}_{12}$, $a=0.75+0.25\tan\beta$
and $\lambda_{12}=\lambda_{21}=0.06$ ($m_{\chi^{\pm}_{1,2}}=101.8\,\rm{GeV}$).
The red region is where the prediction for $\Omega_{\chi} h^2$ is consistent
with the measured one $\sigma$ range of $\Omega_{\mathrm{CDM}}h^2 = 0.1099 \pm 0.0062$.
The dark green region corresponds to $D<3$ ($D$ is defined in subsection 6.2) while
the pale green region represents the part of the parameter space in which $D$ varies
from $3$ to $4$. The grey area indicates that $D>4$. The blue region corresponds to
$m_{\chi^0_1}>M_{Z}/2$, while the dark blue region to the right is ruled out by the
requirement that perturbation theory remains valid up to the GUT scale.}
\label{fig1}
\end{figure}

In order to illustrate the features of the E$_6$SSM mentioned in the previous section,
we shall specify a set of benchmark points (see Tables 1-2).
For each benchmark scenario we calculate the spectrum of the Inert neutralinos, Inert charginos
and Higgs bosons as well as their couplings, the branching ratios of the decays of the lightest CP-even
Higgs state and the dark matter relic density. In order to calculate the dark matter relic density
we use numerical methods. In particular, {\tt MicrOMEGAs 2.2} \cite{MicrOMEGA} is used to numerically compute the
present day density of dark matter. This includes the relevant (co-)annihilation channel cross sections and
the LSP freeze-out temperature. {\tt MicrOMEGAs} achieves this by calculating all of the relevant tree-level
Feynman diagrams using {\tt CalcHEP}. The {\tt CalcHEP} model files for the considered model are generated using
{\tt LanHEP} \cite{LanHEP}. The {\tt MicrOMEGAs} relic density calculation assumes standard cosmology in which
the LSP was in equilibrium with the photon at some time in the past.

\begin{figure}[t]
\begin{center}
\includegraphics[width=150mm]{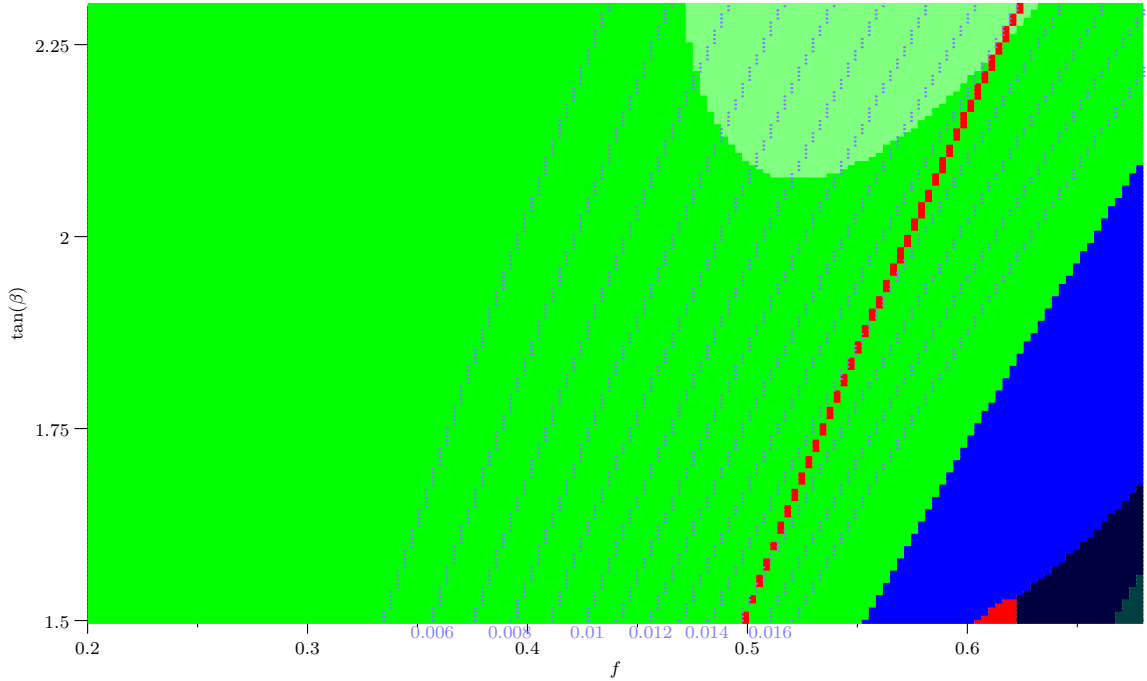}
\end{center}
\caption{\it\small Contour plot of $(X^{h_1}_{11})^2$ and relic density
$\Omega_\chi h^2$ regions in the $(f,\tan \beta)$-plane with $s=2400\,\rm{GeV}$,
$f_{\alpha\alpha}=\tilde{f}_{\alpha\alpha}=\lambda_{\alpha\alpha}=0$,
$f_{12}=f$, $\tilde{f}_{12}=f_{12}/a$, $f_{21}=1.02\cdot f_{12}$,
$\tilde{f}_{21}=0.98\cdot \tilde{f}_{12}$, $a=0.5+0.5\tan\beta$
and $\lambda_{12}=\lambda_{21}=0.06$ ($m_{\chi^{\pm}_{1,2}}=101.8\,\rm{GeV}$).
The red region is where the prediction for $\Omega_{\chi} h^2$ is consistent with
the measured one $\sigma$ range of $\Omega_{\mathrm{CDM}}h^2 = 0.1099 \pm 0.0062$.
The dark green region corresponds to $D<3$ while the pale green region represents
the part of the parameter space in which $D$ varies from $3$ to $4$.
The blue region corresponds to $m_{\chi^0_1}>M_{Z}/2$, while the dark blue region
to the right is ruled out by the requirement that perturbation theory remains
valid up to the GUT scale.}
\label{fig2}
\end{figure}

\subsection{Benchmark scenarios}



In order to construct benchmark scenarios that are consistent with cosmological observations
and collider constraints we restrict our considerations to low values of $\tan\beta\lesssim 2$.
Figs.~1 and 2 show that in principle the appropriate value of dark matter density can be obtained
even when $\tan\beta > 2$. At the same time larger values of $\tan\beta$ lead to masses of the
lightest and second lightest Inert neutralinos that are too small, as discussed in Section 3.
As a result larger couplings of the lightest Inert neutralinos to $Z$ are required to
reproduce the measured value of $\Omega_{\mathrm{CDM}}h^2$. On the other hand
according to Figs.~1 and 2, light Inert neutralinos with substantial couplings to $Z$--boson
give a considerable contribution to its invisible width leading to a conflict with LEP
measurements (see discussion in subsection 6.2).

However, even for $\tan\beta\lesssim 2$ the lightest inert neutralino states can get
appreciable masses only if either all or at least one of the Inert chargino mass eigenstates
are light, i.e. $m_{\chi^{\pm}_1}\simeq 100-200\,\mbox{GeV}$. As clarified in
Section 3 and in \cite{Hall:2009aj}, the masses of the lightest
inert neutralino states decrease with increasing $m_{\chi^{\pm}_{1,\,2}}$ and it is therefore
rather difficult to find benchmark scenarios consistent with cosmological observations for
$m_{\chi^{\pm}_{1}}\gtrsim 200\,\mbox{GeV}$. At the same time we demonstrate (see benchmark
point (ix) in Table 2) that one light Inert chargino mass eigenstate is enough to ensure
that the lightest inert neutralino state gains a mass of the order of $M_Z/2$.

To obtain the kind of Inert neutralino and chargino spectrum discussed above one has to
assume that some or all of the couplings $\lambda_{\alpha\beta}$ are rather small, e.g. they are expected
to be much smaller than $f_{\alpha\beta}$ and $\tilde{f}_{\alpha\beta}$. On the other hand in order
to get $m_{\chi_1^0}\sim m_{\chi_2^0}\sim M_Z/2$ the Yukawa couplings $f_{\alpha\beta}$ and
$\tilde{f}_{\alpha\beta}$ need to be relatively close to their theoretical upper bounds which are
caused by the requirement of the validity of perturbation theory up to the GUT scale.
Since gauge coupling unification determines the RG flow and low energy value of $g'_1$ the mass
of the $Z'$ gauge boson is set by the VEV of the singlet field $s$ only. In our study we choose
$s=2400\,\mbox{GeV}$ so that the $Z'$ mass is about $890\,\mbox{GeV}$. This value of the $Z'$ boson
mass is just above the present lower bound of $865\,\mbox{GeV}$ set by CDF \cite{Erler:2010uy}
and allows satisfaction of stringent limits on the $Z'$ mass and $Z-Z'$ mixing that come from
precision EW tests \cite{Erler:2009jh}.

Since we restrict our analysis to low values of $\tan\beta\lesssim 2$ the mass of the SM-like
Higgs boson is very sensitive to the choice of the coupling $\lambda$. Stringent LEP constraints
require $\lambda(M_t)$ to be larger than the low energy value of $g'_1\simeq 0.47$. If we try to
increase $\lambda(M_t)$ much further, then the theoretical upper bounds on $f_{\alpha\beta}$ and
$\tilde{f}_{\alpha\beta}$ become substantially stronger. As a consequence, it is rather difficult
to find solutions with $|m_{\chi_1^0}|\sim |m_{\chi_2^0}|\sim M_Z/2$. Therefore in our analysis we
concentrate on values of $\lambda(M_t)\lesssim 0.6$. In addition, we set stop scalar masses
to be equal to $m_Q=m_U=M_S=700\,\mbox{GeV}$ and restrict our consideration to the so-called
maximal mixing scenario when the stop mixing parameter $X_t=A_t-\lambda s/(\sqrt{2}\tan\beta)$
is equal to $X_t=\sqrt{6} M_S$. This choice of parameters limits the range of variations of
the lightest CP--even Higgs mass. In the leading two--loop approximation the mass of the
SM-like Higgs boson varies from $115\,\mbox{GeV}\,(\lambda=g'_1)$ to $136\,\mbox{GeV}\,(\lambda=0.6)$.
From Tables 1-2 one can see that the large values of $\lambda\gtrsim g'_1$ that
we choose in our analysis result in the extremely hierarchical structure of the Higgs spectrum,
as pointed out in Section 4 (see also \cite{King:2005jy}). In Tables 1-2 the masses of the heavy 
Higgs states are computed in the leading one--loop approximation. In the case of the lightest
Higgs boson mass the leading two--loop corrections are taken into account.

The set of the benchmark points that we specify demonstrates that one can get a reasonable
dark matter density consistent with the recent observations if $|m_{\chi_1^0}|\sim |m_{\chi_2^0}|\sim M_Z/2$.
Our benchmark scenarios also indicate that in this case the SM--like Higgs boson decays predominantly
into the lightest inert neutralinos ($\chi_1$ and $\chi_2$) while the total branching ratio
into SM particles varies from 2\% to 4\%.

The benchmark points (i), (ii), (iv), (v) and (viii) are motivated by a non-Abelian family
symmetry $\Delta_{27}$ which describes well the observed hierarchy in the quark and lepton sectors.
As was discussed in Section 3 these scenarios imply that all flavour diagonal Yukawa couplings
$\lambda_{\alpha\alpha}$, $f_{\alpha\alpha}$ and $\tilde{f}_{\alpha\alpha}$ are rather small.
Due to the approximate global $U(1)$ symmetry (\ref{icn19}), that originates from the family symmetry
$\Delta_{27}$, the spectrum of Inert neutralinos involves a set of pseudo--Dirac states.
When the masses of the lightest and second lightest Inert neutralinos are close or they form a
Dirac state then the decays of $h_1$ into $\chi_{\alpha}\chi_{\beta}$ will not be observed at the
LHC. Thus these decay channels give rise to a large invisible branching ratio of the SM--like Higgs boson.

In Tables 1-2 we presented a few benchmark scenarios (i), (ii), (iv)-(vi), (ix) with almost degenerate
lightest and second lightest Inert neutralinos. In some of these benchmark points both lightest Inert
neutralinos are lighter than $M_Z/2$. Thus the $Z$--boson can decay into $\chi_{\alpha}\chi_{\beta}$ so that
the lightest and second lightest Inert neutralino states contribute to the invisible $Z$--boson width.
In other benchmark scenarios both of the lightest Inert neutralinos have masses above $M_Z/2$ and
the decays $Z\to\chi_{\alpha}\chi_{\beta}$ are kinematically forbidden.

When the LSP and NLSP are close in mass, LSP-NLSP co-annihilations may be an important factor in 
determining the dark matter relic density. If this is the case then the LSP-NLSP mass splitting should 
be an important factor. Since annihilations of two like-neutralinos are p-wave suppressed, one should 
compare $\beta R_{Z11}$
with $R_{Z12}$ when trying to determine how important co-annihilations are,
where $\beta$ is the relative speed of the incoming particles, approximately 1/6.
It is useful to consider the following situations. With the LSP and NLSP almost degenerate
and with equal self-annihilation cross-sections, but a negligible co-annihilation cross-section,
the relic density of dark matter would be twice what it would have have been if the NLSP
had not been present. If, alternatively, the co-annihilation cross-section was
equal to the self-annihilation cross-sections, the existence of this extra channel would lead to a lower
relic density. In this case it would in fact be equal to the relic density
calculated in the absence of the NLSP. In this way, in such a scenario where co-annihilations
and self-annihilations are about as important as each other, the relic density is largely independent
of the LSP-NLSP mass splitting.

For the benchmark scenarios (i) and (ii) this latter situation is approximately the case
and the LSP-NLSP mass splitting turns out not to be an important
factor. The mass splitting is in fact small, about half a GeV, but if it were larger
and the NLSPs were made to have frozen-out much earlier, the relic density would only
be decreased slightly (by about a tenth).
In benchmark scenario (iv), even though the LSP and NLSP are close in mass, co-annihilations 
are unimportant due to the small value of $R_{Z12}$. In this case increasing the NLSP mass substantially 
while keeping everything else fixed would lead to an approximate halving of the predicted relic density, 
since the NLSPs would have decoupled much earlier than, rather than at the same time as, the LSPs. 
The only other benchmark scenario where the LSP and NLSP are close enough in mass for co-annihilations 
to be potentially important is scenario (ix). Here co-annihilation is in fact the dominant process and 
changing the LSP-NLSP mass splitting would have a large effect on the predicted relic density. In fact, 
in this scenario, if the NLSP were not present the predicted relic density would be within 
the measured range.

If the mass difference between the second lightest and the lightest Inert neutralino is
$10\,\mbox{GeV}$ or more, then some of the decay products of a $\chi_2$ that originates from a
SM-like Higgs boson decay might be observed at the LHC. In our analysis we assume that
all scalar particles, except for the lightest Higgs boson, are heavy and that the couplings of the
Inert neutralino states to quarks, leptons and their superpartners are relatively small.
As a result the second lightest Inert neutralino decays into the lightest one and a
fermion--antifermion pair mainly via a virtual $Z$. In our numerical analysis we did
not manage to find any benchmark scenario with $|m_{\chi_2^0}|-|m_{\chi_1^0}|\gtrsim 20\,\mbox{GeV}$
leading to reasonable values of $\Omega_{\mathrm{CDM}}h^2$. Hence we do not expect
any observable jets at the LHC associated with the decay of a
$\chi_2$ produced through a Higgs decay. However, it might be possible to detect some
lepton-antilepton pairs that come from the decays $h_1\to\chi_{2}\chi_{\alpha}$.
In particular, we hope that $\mu^{+} \mu^{-}$ pairs that come from the exotic
decays of the lightest CP--even Higgs state mentioned above can be observed at the LHC.

In Tables 1-2 benchmark scenarios (iii), (vii), (viii) can lead to these relatively
energetic muon pairs in the final state of the SM-like Higgs decays. Since the Higgs branching
ratios into SM particles are rather suppressed, the decays of the
lightest CP--even Higgs state into $l^{+} l^{-} + X$ might play an essential role in
Higgs searches.

In addition to the exotic Higgs decays, the scenarios considered here imply that
at least two of the Inert neutralino states that are predominantly
the fermion components of the Inert Higgs doublet superfields and one of the Inert chargino states
should have masses below
$200\,\mbox{GeV}$. Since these states are almost Inert Higgsinos they couple rather
strongly to $W$ and $Z$--bosons. Thus at hadron colliders the corresponding Inert
neutralino and chargino states can be produced in pairs via off-shell $W$ and $Z$--bosons.
Since they are light their production cross sections at the LHC are not negligibly
small. After being produced Inert neutralino and chargino states sequentially decay into
the LSP and pairs of leptons and quarks resulting in distinct signatures that can be
discovered at the LHC in the near future.

\subsection{Neutralino and chargino collider limits}

The remarkable signatures discussed above raise serious concerns that they could have
already been observed at the Tevatron and/or even earlier at LEP. For example, the light Inert
neutralino and chargino states could be produced at the Tevatron \cite{Baer:1992dc}.
Recently, the CDF and D0 collaborations set a stringent lower bound on chargino
masses using searches for SUSY with a trilepton final state (i.e. trilepton signal) \cite{trilepton}.
These searches ruled out chargino masses below $164\,\mbox{GeV}$. However this lower
bound on the chargino mass was obtained by assuming that the corresponding chargino
and neutralino states decay predominantly into the LSP and a pair of leptons. In our case,
however, the Inert neutralino and chargino states are expected to decay
via virtual $Z$ and $W$ exchange, i.e. they decay predominantly into the LSP and
a pair of quarks. As a consequence the lower limit on the mass of charginos that is set
by the Tevatron is not directly applicable to the benchmark scenarios that we consider
here. Instead in our study we use the $95\%\,\mbox{C.L.}$ lower limit on the chargino mass
of about $100\,\mbox{GeV}$ that was set by LEP II \cite{Kraan:2005vy}.

In principle LEP experiments also set constraints on the masses and couplings of
neutral particles that interact with the $Z$--boson.
As mentioned above when the masses of $\chi_1$ and $\chi_2$ are below $M_Z/2$
they are almost degenerate and thus the decays of $Z$ into $\chi_{\alpha}\chi_{\beta}$
contribute to the invisible width of the $Z$--boson changing the effective number of
neutrino species $N_{\nu}^{eff}$. The contribution of $\chi_1$ and $\chi_2$
($\Delta N_{\nu}^{eff}$) to $N_{\nu}^{eff}$ is given by
\be
\Delta N_{\nu}^{eff}=\delta_{11}+2\delta_{12}+\delta_{22}\,,
\label{higgs14}
\ee
where
\be
\ba{rcl}
\delta_{\alpha\beta}&= &R_{Z\alpha\beta}^2\biggl[1-\dfrac{|m_{\chi^0_{\alpha}}|^2+|m_{\chi^0_{\beta}}|^2}{2 M_Z^2}
-3(-1)^{\theta_{\alpha}+\theta_{\beta}}\dfrac{|m_{\chi^0_{\alpha}}||m_{\chi^0_{\beta}}|}{M_Z^2}\\[4mm]
&-&\dfrac{(|m_{\chi^0_{\alpha}}|^2-|m_{\chi^0_{\beta}}|^2)^2}{2 M_Z^4}\biggr]\sqrt{\biggl(1-\dfrac{|m_{\chi^0_{\alpha}}|^2+|m_{\chi^0_{\beta}}|^2}{M_Z^2}\biggr)^2
-4\dfrac{|m_{\chi^0_{\alpha}}|^2 |m_{\chi^0_{\beta}}|^2}{M_Z^4}}\,.
\ea
\label{higgs15}
\ee
All three terms in Eq.~(\ref{higgs14}) contribute to $N_{\nu}^{eff}$ only if $2|m_{\chi^0_{2}}|<M_Z$.
In the case where only the $Z$--boson decays into $\chi^0_1\chi^0_1$ are kinematically allowed the values
of $\delta_{12}$ and $\delta_{22}$ should be set to zero. If $|m_{\chi^0_{1}}|+|m_{\chi^0_{2}}|<M_Z$
while $2|m_{\chi^0_{2}}|>M_Z$ then only $\delta_{11}$ and $\delta_{12}$ need to be taken into account.

In order to compare the measured value of $N_{\nu}$ with the effective number of neutrino species
in the E$_6$SSM, i.e. $N_{\nu}^{eff}=3+\Delta N_{\nu}^{eff}$, it is convenient to define the variable
\be
D=\dfrac{N_{\nu}^{eff}-N_{\nu}^{exp}}{\sigma^{exp}}\,,
\label{higgs16}
\ee
where $N_{\nu}^{exp}=2.984$ and $\sigma^{exp}=0.008$ \cite{pdg}. The value of $D$ represents the deviation
between the predicted and measured effective number of neutrinos contributing to the $Z$--boson invisible
width. It is worth pointing out that in the SM $D=2$. In the benchmark scenarios presented in Tables 1-2
the value of $D$ is always less than 3. Figs.~1 and 2 also demonstrate that there is a substantial part
of the E$_6$SSM parameter space where $m_{\chi^0_{1,2}}<M_Z/2$ and $D<3$.
This indicates that the relatively light Inert neutralinos
with masses below $M_Z/2$ are not ruled out by different constraints on the effective number
of neutrinos set by LEP experiments (see, for example \cite{pdg}--\cite{Z-inv-width}).
Indeed, as argued in Section 3 the Yukawa couplings $f_{\alpha\beta}$ and
$\tilde{f}_{\alpha\beta}$ can be chosen such that the $R_{Z\alpha\beta}$ are very small.
The couplings of the lightest and second lightest Inert neutralinos to the $Z$--boson
are relatively small anyway because of the Inert singlino admixture in these states.
Nevertheless Figs.~1 and 2 show that the scenarios with light Inert neutralinos which have
masses below $M_Z/2$ and relatively small couplings to the $Z$--boson can lead to the 
appropriate dark matter density consistent with the recent observations.

LEP has set limits on the cross section of $e^{+}e^{-}\to\chi_2^0\chi_1^0\,(\chi_1^{+}\chi_1^{-})$
in the case when $\chi_2^0\to q\bar{q}\chi^0_1\,(\chi_1^{\pm}\to q\bar{q}'\chi^0_1)$ predominantly
\cite{Abbiendi:2003sc}. Unfortunately, the bounds are not directly applicable for our study because
OPAL limits were set for a relatively heavy $\chi_2^0\, (\chi_1^{\pm})$ only
($|m_{\chi^0_2}|\gtrsim 60\,\mbox{GeV}$). Nevertheless, these bounds demonstrate that
it was difficult to observe light neutralinos with $|m_{\chi^0_{2,1}}|\lesssim 100\,\mbox{GeV}$ if
their production cross section
$\sigma(e^{+}e^{-}\to\chi_{\alpha}^0\chi_{\beta}^0)\lesssim 0.1-0.3\,\mbox{pb}^{-1}$.
Since at LEP energies the cross sections of colourless particle
production through s-channel $\gamma/Z$ exchange are typically a few picobarns the
lightest and second lightest Inert neutralino states in the E$_6$SSM could escape detection at
LEP if their couplings $R_{Z\alpha\beta}\lesssim 0.1-0.3$.

\subsection{Dark matter direct detection}

Another constraint on the couplings of the lightest Inert neutralino comes from experiments
for the direct detection of dark matter. Recently the CDMSII and XENON100 collaborations have set
upper limits on the weakly interacting massive particle (WIMP)--nucleon elastic--scattering
spin--independent cross section \cite{Ahmed:2009zw},\cite{Aprile:2010um}.  The XENON100 Collaboration claims a limit on the spin-independent cross section of $3.4\times 10^{-44}$~cm$^2$ for a 55~GeV WIMP. This limit remains fairly constant for lower WIMP masses and does not increase above about $4\times 10^{-44}$~cm$^2$ even for the lowest LSP masses that are consistent with our thermal freeze-out scenario.  Since in the E$_6$SSM
the couplings of the lightest Inert neutralino to quarks (leptons) and squarks (sleptons) are
suppressed, the $\chi_1^0$--nucleon elastic scattering, which is associated with the spin-independent
cross section, is mediated mainly by the $t$--channel lightest Higgs boson exchange. Thus in
the leading approximation the spin--independent part of $\chi_1^0$--nucleon cross section in the
E$_6$SSM takes the form \cite{Ellis:2008hf,Kalinowski:2008iq}
\be
\ba{l}
\sigma_{SI}=\dfrac{4 m^2_r m_N^2}{\pi v^2 m^4_{h_1}} |X^{h_1}_{11} F^N|^2\,,\\[4mm]
m_r=\dfrac{m_{\chi^0_1} m_N}{m_{\chi^0_1}+m_N}\,,\qquad\qquad\qquad
F^N=\sum_{q=u,d,s} f^N_{Tq} + \dfrac{2}{27}\sum_{Q=c,b,t} f^N_{TQ}\,,
\ea
\label{higgs17}
\ee
where
$$
m_N f^N_{Tq} = \langle N | m_{q}\bar{q}q |N \rangle\,, \qquad\qquad\qquad\qquad
f^N_{TQ} = 1 - \sum_{q=u,d,s} f^N_{Tq}\,.
$$
Here for simplicity we assume that the lightest Higgs state has the same couplings
as the Higgs boson in the SM and ignore all contributions induced by heavy Higgs
and squark exchange\footnote{The presence of almost degenerate lightest and second 
lightest Inert neutralinos could result in the inelastic scattering of $\chi_1^0$
on nuclei ($A$), i.e. $\chi_1^0+A\to \chi_2^0+A$, that could affect the direct
detection of $\chi_1^0$ at the experiment. However such processes may take place
only if the mass splitting between $\chi_1^0$ and $\chi_2^0$ is less than 
$100\,\mbox{KeV}$ \cite{TuckerSmith:2001hy}. Since in all of the benchmark scenarios
considered here the corresponding mass splitting is substantially larger the
inelastic scattering of $\chi_1^0$ does not play any significant role.}. 
Due to the hierarchical structure of the particle spectrum and the approximate 
$Z_2^H$ symmetry this approximation works very well. Using the experimental limits 
set on $\sigma_{SI}$ and Eqs.~(\ref{higgs17}) one can obtain upper bounds 
on $X^{h_1}_{11}$ \cite{Cheung:2009wb}.

In Tables 1-2 we specify the interval of variations of $\sigma_{SI}$ for each benchmark
scenario. As one can see from Eq.~(\ref{higgs17}) the value of $\sigma_{SI}$ depends
rather strongly on the hadronic matrix elements, i.e. the coefficients $f^N_{Tq}$, that
are related to the $\pi$--nucleon $\sigma$ term and the spin content of the nucleon.
The hadronic uncertainties in the elastic scattering cross section of dark matter
particles on nucleons were considered in \cite{Ellis:2008hf,Bottino:1999ei}.
In particular, it was pointed out that $f^N_{Ts}$ could vary over a wide range.
In Tables 1-2 the lower limit on $\sigma_{SI}$ corresponds to $f^N_{Ts}=0$ while the upper
limit implies that $f^N_{Ts}=0.36$ (see \cite{Kalinowski:2008iq}). From Tables 1-2 and
Eq.~(\ref{higgs17}) it also becomes clear that $\sigma_{SI}$ decreases substantially
when $m_{h_1}$ grows.

Since in all of the benchmark scenarios presented in Tables 1-2
the lightest Inert neutralino is relatively heavy $(|m_{\chi^0_{1}}|\sim M_Z/2)$,
allowing for a small enough dark matter relic density,
the coupling of $\chi_1^0$ to the lightest CP-even Higgs state is always large giving
rise to a $\chi_1^0$--nucleon spin-independent cross section which is of the order of
or  larger than the experimental upper bound. However it is worth keeping in
mind that the obtained experimental limits on $\sigma_{SI}$ are not very robust
\cite{Collar:2010ht}. Moreover, CDMS II and XENON100 quote $90\%$ C.L. upper bounds
while the $95\%$ confidence level bounds are larger by a factor of 1.3\,.
By the same token the $99\%$ C.L. and $99.9\%$ C.L. upper bounds, which are associated
with $2.6$ and $3.3$ standard deviations, are expected to be $2$ and $3$ times larger
than the $90\%$ C.L. bounds respectively. Following these estimates it is clear that
the benchmark scenarios presented in Tables 1-2 cannot yet be ruled out by either XENON100 or
CDMS II. However in the near future the expected new analysis from XENON100
may either confirm or refute our scenario.

\section{Summary and Conclusions}

In this paper we have considered novel decays of the SM--like Higgs boson
which can occur within a particular dark matter motivated scenario of the
Exceptional Supersymmetric Standard Model (E$_6$SSM). This model implies that
at high energies the $E_6$ GUT gauge group is broken to the SM gauge group together
with an additional $U(1)_N$ gauge group under which right--handed neutrinos have zero
charge. To ensure anomaly cancellation and gauge coupling unification, the low energy
matter content of the E$_6$SSM includes three $27$ representations of $E_6$ and a pair
of $SU(2)$ doublets from an additional $27'$ and $\overline{27'}$. Thus the E$_6$SSM
involves extra exotic matter beyond that of the MSSM that includes two families of Inert Higgs
doublet superfields $H^u_{\alpha}$ and $H^d_{\alpha}$ and two Inert SM singlet superfields
$S_{\alpha}$ that carry $U(1)_{N}$ charges. The fermion components of these superfields
form Inert neutralino and chargino states.

To satisfy LEP constraints we restricted our consideration to scenarios with relatively
heavy Inert chargino states, i.e. $m_{\chi^{\pm}_{1,2}}\gtrsim 100\,\mbox{GeV}$. In our
analysis we also required the validity of perturbation theory up to the GUT scale which
sets stringent constraints on the values of the Yukawa couplings at low energies.
Using these restrictions we argued that the lightest and the second lightest Inert
neutralinos ($\chi_1^0$ and $\chi_2^0$) are always light, viz. they typically have masses
below $60-65\,\mbox{GeV}$. These neutralinos are mixtures of Inert Higgsinos and singlinos. In our
model $\chi_1^0$ tends to be the LSP  and can play the
role of dark matter, while $\chi_2^0$ tends to be the NLSP.
The masses of $\chi_1^0$ and $\chi_2^0$ can be induced even if only one
family of the Inert Higgsinos couples to the two SM singlinos. The masses of $\chi_1^0$ and $\chi_2^0$
decrease with increasing $\tan\beta$ and Inert chargino masses.

An important requirement of this paper is that the lightest Inert neutralino account for all or most of
the observed dark matter relic density. This sets another stringent constraint on the masses
and couplings of $\chi_1^0$. Indeed, because the lightest Inert neutralino states are almost
Inert singlinos, their couplings to the gauge bosons, Higgs states, quarks (squarks)
and leptons (sleptons) are rather small resulting in a relatively small annihilation
cross section of $\tilde{\chi}^0_1\tilde{\chi}^0_1\to \mbox{SM particles}$ and the possibility
of an unacceptably large dark matter density. In the limit when all non-SM states except the
Inert neutralinos and charginos are heavy ($\gtrsim \mbox{TeV}$) a reasonable density of dark
matter can be obtained for $|m_{\chi^0_{1,\,2}}|\sim M_Z/2$
where the Inert LSPs annihilate mainly through $Z$ in the $s$--channel \cite{Hall:2009aj}.
If $\tilde{\chi}^0_1$
annihilation proceeds through the $Z$--boson resonance, i.e. $2|m_{\chi^0_{\sigma}}|\approx M_Z$,
then an appropriate value of $\Omega_{\mathrm{CDM}}h^2$ can be achieved even for a relatively small
coupling of $\tilde{\chi}^0_1$ to $Z$.

The above scenario naturally emerges when a $\Delta_{27}$ family symmetry is included
in the E$_6$SSM  \cite{Howl:2009ds}. The family symmetry was not introduced for this purpose,
instead it was introduced earlier to provide an explanation of the $Z_2^H$ symmetry and to
account for the quark and lepton masses and mixings, including tri-bimaximal neutrino mixing.
It is therefore encouraging to find that the same symmetry leads to a spectrum of inert
pseudo-Dirac neutralinos which allows for a successful dark matter relic abundance, and
also predicts novel Higgs decays. The $\Delta_{27}$ family symmetry also implies
two almost degenerate families of $D$--fermion states \cite{Howl:2009ds} and in addition
may have interesting consequences for $B$--physics \cite{King:2010np}.
As discussed in subsection (3.2) this symmetry leads to a cancellation
of different contributions to the off-diagonal couplings of the LSP and NLSP.
In addition, due to the singlino component of the lightest Inert neutralino states,
the diagonal couplings of $\chi_1^0$ and $\chi_2^0$ to the $Z$--boson can also be rather small. Therefore
these states could have escaped detection at LEP.

The main point we make in this paper is that, within the above dark matter motivated scenario,
although the lightest and the second lightest Inert neutralinos might have very small couplings to
the $Z$--boson, their couplings to the SM--like Higgs state $h_1$ are always large. Indeed, we argued
that in the first approximation the couplings of $\chi_1^0$ and $\chi_2^0$ to the lightest CP--even
Higgs boson are proportional to $|m_{\chi^0_{1,\,2}}|/v$. Since $|m_{\chi^0_{1,\,2}}|\sim M_Z/2$
these couplings are much larger than the corresponding $b$--quark coupling. Thus the SM--like Higgs
boson decays predominantly into the lightest inert neutralino states and has very small branching
ratios ($2\%-4\%$) for decays into SM particles. We have illustrated this, together with the other
phenomenological aspects of the dark matter motivated scenario considered in this paper,
by presenting a set of benchmark points in Tables 1-2. If the masses of the
lightest and second lightest Inert neutralinos are very close then the decays of
$h_1$ into $\chi_{\alpha}\chi_{\beta}$ will not be observed at the LHC giving rise to a large invisible
branching ratio of the SM--like Higgs boson. When the mass difference between the second lightest and
the lightest Inert neutralinos is larger than $10\,\mbox{GeV}$ the invisible branching ratio
remains dominant but some of the decay products of $\chi_2$ might be observed at the LHC.
In particular, there is a chance that  $\mu^{+} \mu^{-}$ pairs could be detected. Since the
branching ratios of $h_1$ into SM particles are extremely suppressed, the decays of the SM--like
Higgs boson into $l^{+} l^{-} + X$ could be important for Higgs searches.


In conclusion, the E$_6$SSM predicts three Higgs families plus three Higgs singlets, where
one family develop VEVs, while the remaining two which do not are called Inert.
This pattern of Higgs VEVs is due to a broken $Z_2^H$ symmetry whose origin can be understood from a
$\Delta_{27}$ family symmetry.
The model can account for the dark matter relic abundance if
the two lightest Inert neutralinos, identified as the LSP and NLSP, have
masses close to half the $Z$ mass, with a pseudo-Dirac structure as predicted by the
$\Delta_{27}$ family symmetry.
Within this scenario we find that the usual SM-like Higgs boson
decays more that 95\% of the time into either LSPs or NLSPs,
with the latter case producing a final state containing two soft leptons $l^{+} l^{-}$ with an invariant mass
less than or about 10 GeV. We have illustrated this with a set of benchmark points satisfying
phenomenological constraints and the WMAP dark matter relic abundance.
This scenario also predicts other light Inert chargino and neutralino states
below $200\,\mbox{GeV}$, and large LSP direct detection cross-sections close
to current limits and observable soon at XENON100.

\section*{Acknowledgements}
\vspace{-2mm} We would like to thank P.~Athron, A.~Belyaev, E.~E.~Boos, M.~Drees, I.~F.~Ginzburg, M.~Krawczyk, 
J.~P.~Kumar, D.~J.~Miller, D.~Melikhov, S.~Moretti, N.~V.~Nikitin, L.~B.~Okun, V.~A.~Rubakov, B.~D.~Thomas, 
D.~G.~Sutherland, I.~P.~Volobuev, M.~I.~Vysotsky for fruitful discussions. The authors are grateful to 
X.~R.~Tata for valuable comments and remarks. The work of R.N. and S.P. was supported by the U.S. Department 
of Energy under Contract DE-FG02-04ER41291, and the work of M.S. was supported by the National Science 
Foundation PHY-0755262. S.F.K. acknowledges partial support from the STFC Rolling Grant ST/G000557/1.
J.P.H. is thankful to the STFC for providing studentship funding.

\renewcommand{\baselinestretch}{1.00}

\begin{table}[ht]
\centering
\begin{tabular}{|c||c|c|c|c|}
\hline
                            &	i	    &	ii	    &	iii	    &	iv	\\\hline\hline
$\tan(\beta)$	            &	1.5	    &	1.5	    &	1.7	    &	1.564	\\\hline
$m_{H^{\pm}}\simeq m_{A}\simeq m_{h_3}$/GeV&1977& 1977&	2022	&	1990	\\\hline
$m_{h_1}$/GeV	            &	135.4	&	135.4	&	133.1	&	134.8	\\\hline\hline
$\lambda_{22}$	            &	0.001	&	0.001	&	0.094	&	0.0001	\\\hline
$\lambda_{21}$	            &	0.077	&	0.062	&	0	    &	0.06	\\\hline
$\lambda_{12}$	            &	0.077	&	0.062	&	0	    &	0.06	\\\hline
$\lambda_{11}$	            &	0.001	&	0.001	&	0.059	&	0.0001	\\\hline\hline
$f_{22}$	                &	0.001	&	0.001	&	0.53	&	0.001	\\\hline
$f_{21}$	                &	0.61	&	0.61	&	0.05	&	0.476	\\\hline
$f_{12}$	                &	0.6	    &	0.6	    &	0.05	&	0.466	\\\hline
$f_{11}$	                &	0.001	&	0.001	&	0.53	&	0.001	\\\hline\hline
$\tilde{f}_{22}$	        &	0.001	&	0.001	&	0.53	&	0.001	\\\hline
$\tilde{f}_{21}$	        &	0.426	&	0.426	&	0.05	&	0.4	\\\hline
$\tilde{f}_{12}$	        &	0.436	&	0.436	&	0.05	&	0.408	\\\hline
$\tilde{f}_{11}$	        &	0.001	&	0.001	&	0.53	&	0.001	\\\hline\hline
$m_{\tilde{\chi}^0_1}$/GeV	&	41.91	&	47.33	&	33.62	&	-36.69	\\\hline
$m_{\tilde{\chi}^0_2}$/GeV	&	-42.31	&	-47.84	&	47.78	&	36.88	\\\hline
$m_{\tilde{\chi}^0_3}$/GeV	&	-129.1	&	-103.6	&	108.0	&	-103.11	\\\hline
$m_{\tilde{\chi}^0_4}$/GeV	&	132.4	&	107.0	&	-152.1	&	103.47	\\\hline
$m_{\tilde{\chi}^0_5}$/GeV	&	171.4	&	151.5	&	163.5	&	139.80	\\\hline
$m_{\tilde{\chi}^0_6}$/GeV	&	-174.4	&	-154.4	&	-200.8	&	-140.35	\\\hline\hline
$m_{\tilde{\chi}^\pm_1}$/GeV&	129.0	&	103.5	&	100.1	&	101.65	\\\hline
$m_{\tilde{\chi}^\pm_2}$/GeV&	132.4	&	106.9	&	159.5	&	101.99	\\\hline\hline
$\Omega_\chi h^2$	        &	0.096	&	0.098	&	0.109	&	0.107	\\\hline\hline
$R_{Z11}$	                &	-0.0250	&	-0.0407	&	-0.144	&	-0.132	\\\hline
$R_{Z12}$	                &	0.0040	&	0.0048	&	0.051	&	0.0043	\\\hline
$R_{Z22}$	                &	-0.0257	&	-0.0429	&	-0.331	&	-0.133	\\\hline\hline
$\Delta N_\nu^{eff}$	    &  0.000090	&	   0	&	0.0068	&	0.0073	\\\hline
$D$	                        &	2.011	&	2.000	&	2.85	&	2.91	\\\hline\hline
$X^{h_1}_{11}$	            &	0.137	&	0.147	&	0.110	&	-0.114	\\\hline
$X^{h_1}_{12}+X^{h_1}_{21}$	&$-1.9\times 10^{-6}$&$-3.4\times 10^{-6}$&	0.0136	&$1.15\times 10^{-6}$\\\hline
$X^{h_1}_{22}$	            &	-0.138	         &	-0.148	          &	0.125	&0.115	\\\hline\hline
$\sigma_{SI}/10^{-44}$ cm$^2$& 2.6-10.5        & 3.0-12.1	        &1.7-7.1&2.0-8.2\\\hline\hline
$\mathrm{Br}(h\rightarrow \tilde{\chi}^0_1 \tilde{\chi}^0_1)$& 49.5\%	           & 49.7\%              & 57.8\% & 49.1\%\\\hline
$\mathrm{Br}(h\rightarrow \tilde{\chi}^0_1 \tilde{\chi}^0_2)$& $7.9\times 10^{-11}$& $2.5\times 10^{-10}$& 0.34\% & 49.2\%\\\hline
$\mathrm{Br}(h\rightarrow \tilde{\chi}^0_2 \tilde{\chi}^0_2)$& 49.0\%              & 48.5\%              & 39.8\% &$3.5\times 10^{-11}$\\\hline
$\mathrm{Br}(h\rightarrow b\bar{b})$                         & 1.36\%              & 1.58\%              & 1.87\% & 1.59\%\\\hline
$\mathrm{Br}(h\rightarrow \tau\bar{\tau})$                   & 0.142\%             & 0.165\%             & 0.196\%& 0.166\%\\\hline\hline
$\Gamma(h\rightarrow \tilde{\chi}^0_1 \tilde{\chi}^0_1)$/MeV & 98.3	               &85.1                 & 81.7   & 82.9\\\hline
$\Gamma^{tot}$/MeV                                           &198.7                &171.1                & 141.2  & 169.0\\\hline
\end{tabular}
\caption{Benchmark scenarios for $m_{h_1}\approx 133-135\,\rm{GeV}$. The branching ratios and decay widths of the 
lightest Higgs boson, the masses of the Higgs states, Inert neutralinos and charginos as well as the couplings of 
$\tilde{\chi}^0_1$ and $\tilde{\chi}^0_2$ are calculated for $s=2400\,\mbox{GeV}$, $\lambda=0.6$, 
$A_{\lambda}=1600\,\mbox{GeV}$, $m_Q=m_U=M_S=700\,\mbox{GeV}$, $X_t=\sqrt{6} M_S$ that correspond to $m_{h_2}\simeq 
M_{Z'}\simeq 890\,\mbox{GeV}$.
}
\end{table}

\begin{table}[ht]
\centering
\begin{tabular}{|c||c|c|c|c|c|}
\hline
                            &	v	    &	vi	    &	vii	    &  viii               & ix\\\hline\hline
$\tan(\beta)$	            &	1.5	    &	1.7	    &	1.5	    & 1.5                 & 1.5\\\hline
$m_{H^{\pm}}\simeq m_{A}\simeq m_{h_3}$/GeV&1145&1165&  1145	& 1145                & 1145\\\hline
$m_{h_1}$/GeV	            &	115.9   &	114.4	&	115.9	& 115.9               & 115.9\\\hline
\hline
$\lambda_{22}$	            &	0.004	&	0.104	&	0.094	& 0.001               & 0.468\\\hline
$\lambda_{21}$	            &	0.084	&	0	    &	0	    & 0.079               & 0.05\\\hline
$\lambda_{12}$	            &	0.084	&	0	    &	0	    & 0.080               & 0.05\\\hline
$\lambda_{11}$	            &	0.004	&	0.09	&	0.059	& 0.001               & 0.08\\\hline\hline
$f_{22}$	                &	0.025	&	0.72	&	0.53	& 0.04                & 0.05\\\hline
$f_{21}$	                &	0.51	&	0.001	&	0.053	& 0.68                & 0.9\\\hline
$f_{12}$	                &	0.5	    &	0.001	&	0.053	& 0.68                & 0.002\\\hline
$f_{11}$	                &	0.025	&	0.7	    &	0.53	& 0.04                & 0.002\\\hline\hline
$\tilde{f}_{22}$	        &	0.025	&	0.472	&	0.53	& 0.04                & 0.002\\\hline
$\tilde{f}_{21}$	        &	0.49	&	0.001	&	0.053	& 0.49                & 0.002\\\hline
$\tilde{f}_{12}$	        &	0.5	    &	0.001	&	0.053	& 0.49                & 0.05\\\hline
$\tilde{f}_{11}$	        &	0.025	&	0.472	&	0.53	& 0.04                & 0.65\\\hline\hline
$m_{\tilde{\chi}^0_1}$/GeV	&	-35.76	&	41.20	&	35.42	& -45.08              & -46.24\\\hline
$m_{\tilde{\chi}^0_2}$/GeV	&	39.63	&	44.21	&	51.77	& 55.34               & 46.60\\\hline
$m_{\tilde{\chi}^0_3}$/GeV	&	-137.8	&	153.1	&	105.3	& -133.3              & 171.1\\\hline
$m_{\tilde{\chi}^0_4}$/GeV	&	151.7	&	176.7	&	-152.7	& 136.9               & -171.4\\\hline
$m_{\tilde{\chi}^0_5}$/GeV	&	173.6	&	-197.3	&	162.0	& 178.4               & 805.4\\\hline
$m_{\tilde{\chi}^0_6}$/GeV	&	-191.3	&	-217.9	&	-201.7	& -192.2              & -805.4\\\hline\hline
$m_{\tilde{\chi}^\pm_1}$/GeV&	135.8	&	152.7	&	100.1	& 133.0               & 125.0\\\hline
$m_{\tilde{\chi}^\pm_2}$/GeV&	149.3	&	176.5	&	159.5	& 136.8               & 805.0\\\hline\hline
$\Omega_\chi h^2$	        &	0.102	&	0.108	&	0.107	& 0.0324              & 0.00005\\\hline\hline
$R_{Z11}$	                &	-0.116	&	-0.0278	&	-0.115	& -0.0217             & -0.0224\\\hline
$R_{Z12}$	                &	0.0037	&  -0.00039	&	-0.045	& -0.0020             & -0.213\\\hline
$R_{Z22}$	                &	-0.118	&	-0.0455	&	-0.288	& -0.0524             & -0.0226\\\hline\hline
$\Delta N_\nu^{eff}$	    &	0.0049	&  0.00009	&	0.0034	& $1.57\times 10^{-6}$& 0\\\hline
$D$	                        &	2.62	&	2.011	&	2.43	& 2.0002              & 2.0\\\hline\hline
$X^{h_1}_{11}$	            &	-0.117	&	0.141	&	0.117	& -0.147              & -0.148\\\hline
$X^{h_1}_{12}+X^{h_1}_{21}$	& -0.000027	& -0.00025	&	-0.0127	&-0.0000140           & -0.000031\\\hline
$X^{h_1}_{22}$	            &	0.130	&	0.147	&	0.141	& 0.174               & 0.149\\\hline\hline
$\sigma_{SI}/10^{-44}$ cm$^2$&3.9-15.7&5.4-21.9	&3.5-14.2       & 6.0-24.4                    & 6.1-25.0\\\hline\hline
$\mathrm{Br}(h\rightarrow \tilde{\chi}^0_1 \tilde{\chi}^0_1)$& 49.6\%            & 53.5\%            &76.3\%&83.4\%             &49.3\%\\\hline
$\mathrm{Br}(h\rightarrow \tilde{\chi}^0_1 \tilde{\chi}^0_2)$&$2.1\times 10^{-8}$&$7.2\times 10^{-7}$&0.26\%&$7.6\times 10^{-9}$&$3.0\times 10^{-8}$\\\hline
$\mathrm{Br}(h\rightarrow \tilde{\chi}^0_2 \tilde{\chi}^0_2)$& 48.4\%            & 44.2\%            &20.3\%&12.3\%             &47.9\%\\\hline
$\mathrm{Br}(h\rightarrow b\bar{b})$                         & 1.87\%            & 2.04\%            &2.83\%&3.95\%             &2.58\%\\\hline
$\mathrm{Br}(h\rightarrow \tau\bar{\tau})$                   & 0.196\%           & 0.21\%            &0.30\%&0.41\%             &0.27\%\\\hline\hline
$\Gamma(h\rightarrow \tilde{\chi}^0_1 \tilde{\chi}^0_1)$/MeV & 61.5	             & 60.1              &62.6  &49.0               &44.4\\\hline
$\Gamma^{tot}$/MeV                                           &124.1              & 112.2             &82.0  &58.8               &90.1\\\hline
\end{tabular}
\caption{Benchmark scenarios for $m_{h_1}\approx 114-116\,\rm{GeV}$. The branching ratios and decay widths of the 
lightest Higgs boson, the masses of the Higgs states, Inert neutralinos and charginos as well as the couplings of 
$\tilde{\chi}^0_1$ and $\tilde{\chi}^0_2$ are calculated for $s=2400\,\mbox{GeV}$, $\lambda=g'_1=0.468$, 
$A_{\lambda}=600\,\mbox{GeV}$, $m_Q=m_U=M_S=700\,\mbox{GeV}$, $X_t=\sqrt{6} M_S$ that correspond to $m_{h_2}\simeq 
M_{Z'}\simeq 890\,\mbox{GeV}$.
}
\end{table}


\begin{thebibliography}{99}

\bibitem{Chang:2008cw}
S.~Chang, R.~Dermisek, J.~F.~Gunion and N.~Weiner,
Ann.\ Rev.\ Nucl.\ Part.\ Sci.\  {\bf 58} (2008) 75
[arXiv:0801.4554 [hep-ph]];
A.~Djouadi and R.~M.~Godbole,
arXiv:0901.2030 [hep-ph];
R.~Dermisek,
Mod.\ Phys.\ Lett.\  A {\bf 24} (2009) 1631
[arXiv:0907.0297 [hep-ph]].

\bibitem{majoron}
R.~E.~Shrock and M.~Suzuki,
Phys.\ Lett.\  B {\bf 110} (1982) 250;
L.~F.~Li, Y.~Liu and L.~Wolfenstein,
Phys.\ Lett.\  B {\bf 159} (1985) 45;
A.~S.~Joshipura and S.~D.~Rindani,
Phys.\ Rev.\ Lett.\  {\bf 69} (1992) 3269;
A.~S.~Joshipura and J.~W.~F.~Valle,
Nucl.\ Phys.\  B {\bf 397} (1993) 105;
T.~Binoth and J.~J.~van der Bij,
Z.\ Phys.\  C {\bf 75} (1997) 17
[arXiv:hep-ph/9608245];
A.~Datta and A.~Raychaudhuri,
Phys.\ Rev.\  D {\bf 57} (1998) 2940
[arXiv:hep-ph/9708444];
C.~P.~Burgess, M.~Pospelov and T.~ter Veldhuis,
Nucl.\ Phys.\  B {\bf 619} (2001) 709
[arXiv:hep-ph/0011335];
M.~Hirsch, J.~C.~Romao, J.~W.~F.~Valle and A.~Villanova del Moral,
Phys.\ Rev.\  D {\bf 73} (2006) 055007
[arXiv:hep-ph/0512257];
B.~Patt and F.~Wilczek,
arXiv:hep-ph/0605188;
J.~J.~van der Bij,
Phys.\ Lett.\  B {\bf 636} (2006) 56
[arXiv:hep-ph/0603082];
D.~G.~Cerdeno, A.~Dedes and T.~E.~J.~Underwood,
JHEP {\bf 0609} (2006) 067
[arXiv:hep-ph/0607157];
V.~Barger, P.~Langacker, M.~McCaskey, M.~J.~Ramsey-Musolf and G.~Shaughnessy,
Phys.\ Rev.\  D {\bf 77} (2008) 035005
[arXiv:0706.4311 [hep-ph]];
H.~Sung Cheon, S.~K.~Kang and C.~S.~Kim,
JCAP {\bf 0805} (2008) 004
[arXiv:0710.2416 [hep-ph]];
A.~Dedes, T.~Figy, S.~Hoche, F.~Krauss and T.~E.~J.~Underwood,
JHEP {\bf 0811} (2008) 036
[arXiv:0807.4666 [hep-ph]];
H.~Sung Cheon, S.~K.~Kang and C.~S.~Kim,
Phys.\ Lett.\  B {\bf 675} (2009) 203
[arXiv:0807.0981 [hep-ph]];
V.~Barger, P.~Langacker, M.~McCaskey, M.~Ramsey-Musolf and G.~Shaughnessy,
Phys.\ Rev.\  D {\bf 79} (2009) 015018
[arXiv:0811.0393 [hep-ph]];
C.~S.~Kim, S.~C.~Park, K.~Wang and G.~Zhu,
Phys.\ Rev.\  D {\bf 81} (2010) 054004
[arXiv:0910.4291 [hep-ph]];
M.~Farina, D.~Pappadopulo and A.~Strumia,
Phys.\ Lett.\  B {\bf 688} (2010) 329
[arXiv:0912.5038 [hep-ph]];
I.~M.~Shoemaker, K.~Petraki and A.~Kusenko,
JHEP {\bf 1009} (2010) 060
[arXiv:1006.5458 [hep-ph]].


\bibitem{Eboli:1994bm}
O.~J.~P.~Eboli, M.~C.~Gonzalez-Garcia, A.~Lopez-Fernandez, S.~F.~Novaes and J.~W.~F.~Valle,
Nucl.\ Phys.\  B {\bf 421} (1994) 65
[arXiv:hep-ph/9312278];
F.~de Campos, O.~J.~P.~Eboli, J.~Rosiek and J.~W.~F.~Valle,
Phys.\ Rev.\  D {\bf 55} (1997) 1316
[arXiv:hep-ph/9601269].


\bibitem{Choudhury:1993hv}
D.~Choudhury and D.~P.~Roy,
Phys.\ Lett.\  B {\bf 322} (1994) 368
[arXiv:hep-ph/9312347].


\bibitem{Martin:1999qf}
S.~P.~Martin and J.~D.~Wells,
Phys.\ Rev.\  D {\bf 60} (1999) 035006
[arXiv:hep-ph/9903259].



\bibitem{hidden-valley}
R.~Schabinger and J.~D.~Wells,
Phys.\ Rev.\  D {\bf 72} (2005) 093007
[arXiv:hep-ph/0509209];
M.~J.~Strassler and K.~M.~Zurek,
Phys.\ Lett.\  B {\bf 661} (2008) 263
[arXiv:hep-ph/0605193];
S.~Gopalakrishna, S.~J.~Lee and J.~D.~Wells,
Phys.\ Lett.\  B {\bf 680} (2009) 88
[arXiv:0904.2007 [hep-ph]];
S.~Gopalakrishna,
AIP Conf.\ Proc.\  {\bf 1200} (2010) 778
[arXiv:0909.5579 [hep-ph]].


\bibitem{fourth-generation}
K.~Belotsky, D.~Fargion, M.~Khlopov, R.~Konoplich and K.~Shibaev,
Phys.\ Rev.\  D {\bf 68} (2003) 054027
[arXiv:hep-ph/0210153].

\bibitem{diffraction}
K.~Belotsky, V.~A.~Khoze, A.~D.~Martin and M.~G.~Ryskin,
Eur.\ Phys.\ J.\  C {\bf 36} (2004) 503
[arXiv:hep-ph/0406037].


\bibitem{higgs-extraD}
N.~Arkani-Hamed, S.~Dimopoulos, G.~R.~Dvali and J.~March-Russell,
Phys.\ Rev.\  D {\bf 65} (2002) 024032
[arXiv:hep-ph/9811448];
G.~F.~Giudice, R.~Rattazzi and J.~D.~Wells,
Nucl.\ Phys.\  B {\bf 595} (2001) 250
[arXiv:hep-ph/0002178];
N.~G.~Deshpande and D.~K.~Ghosh,
Phys.\ Lett.\  B {\bf 567} (2003) 235
[arXiv:hep-ph/0303160];
A.~Datta, K.~Huitu, J.~Laamanen and B.~Mukhopadhyaya,
Phys.\ Rev.\  D {\bf 70} (2004) 075003
[arXiv:hep-ph/0404056];
D.~Dominici and J.~F.~Gunion,
Phys.\ Rev.\  D {\bf 80} (2009) 115006
[arXiv:0902.1512 [hep-ph]].


\bibitem{Battaglia:2004js}
M.~Battaglia, D.~Dominici, J.~F.~Gunion and J.~D.~Wells,
arXiv:hep-ph/0402062.

\bibitem{Asano:2006nr}
M.~Asano, S.~Matsumoto, N.~Okada and Y.~Okada,
Phys.\ Rev.\  D {\bf 75} (2007) 063506
[arXiv:hep-ph/0602157];
R.~S.~Hundi, B.~Mukhopadhyaya and A.~Nyffeler,
Phys.\ Lett.\  B {\bf 649} (2007) 280
[arXiv:hep-ph/0611116];
L.~Wang and J.~M.~Yang,
Phys.\ Rev.\  D {\bf 79} (2009) 055013
[arXiv:0812.4609 [hep-ph]];


\bibitem{Kanemura:2010sh}
S.~Kanemura, S.~Matsumoto, T.~Nabeshima and N.~Okada,
Phys.\ Rev.\  D {\bf 82} (2010) 055026
[arXiv:1005.5651 [hep-ph]].


\bibitem{Bertone:2004pz}
G.~Bertone, D.~Hooper and J.~Silk,
Phys.\ Rept.\  {\bf 405} (2005) 279
[arXiv:hep-ph/0404175].


\bibitem{Baer:1987eb}
H.~Baer, D.~Dicus, M.~Drees and X.~Tata,
Phys.\ Rev.\  D {\bf 36} (1987) 1363;
J.~F.~Gunion and H.~E.~Haber,
Nucl.\ Phys.\  B {\bf 307} (1988) 445
[Erratum-ibid.\  B {\bf 402} (1993) 569];
K.~Griest and H.~E.~Haber,
Phys.\ Rev.\  D {\bf 37} (1988) 719;
J.~L.~Lopez, D.~V.~Nanopoulos, H.~Pois, X.~Wang and A.~Zichichi,
Phys.\ Rev.\  D {\bf 48} (1993) 4062
[arXiv:hep-ph/9303231];
A.~Djouadi, J.~Kalinowski and P.~M.~Zerwas,
Z.\ Phys.\  C {\bf 57} (1993) 569;
A.~Djouadi, P.~Janot, J.~Kalinowski and P.~M.~Zerwas,
Phys.\ Lett.\  B {\bf 376} (1996) 220
[arXiv:hep-ph/9603368];
G.~Belanger, F.~Boudjema, F.~Donato, R.~Godbole and S.~Rosier-Lees,
Nucl.\ Phys.\  B {\bf 581} (2000) 3
[arXiv:hep-ph/0002039];
G.~Belanger, F.~Boudjema, A.~Cottrant, R.~M.~Godbole and A.~Semenov,
Phys.\ Lett.\  B {\bf 519} (2001) 93
[arXiv:hep-ph/0106275].


\bibitem{Ahmed:2009zw}
Z.~Ahmed {\it et al.}  [The CDMS-II Collaboration],
Science {\bf 327} (2010) 1619
[arXiv:0912.3592 [astro-ph.CO]].



\bibitem{Djouadi:1994mr}
A.~Djouadi,
Int.\ J.\ Mod.\ Phys.\  A {\bf 10} (1995) 1
[arXiv:hep-ph/9406430].

\bibitem{:2001xz}
[LEP Higgs Working for Higgs boson searches and ALEPH Collaboration
and DELPHI Collaboration and CERN-L3 Collaboration and OPAL Collaboration],
arXiv:hep-ex/0107032.



\bibitem{Godbole:2003it}
S.~G.~Frederiksen, N.~Johnson, G.~L.~Kane and J.~Reid,
Phys.\ Rev.\  D {\bf 50} (1994) 4244;
R.~M.~Godbole, M.~Guchait, K.~Mazumdar, S.~Moretti and D.~P.~Roy,
Phys.\ Lett.\  B {\bf 571} (2003) 184
[arXiv:hep-ph/0304137].


\bibitem{Davoudiasl:2004aj}
H.~Davoudiasl, T.~Han and H.~E.~Logan,
Phys.\ Rev.\  D {\bf 71} (2005) 115007
[arXiv:hep-ph/0412269].


\bibitem{Gunion:1993jf}
J.~F.~Gunion,
Phys.\ Rev.\ Lett.\  {\bf 72} (1994) 199
[arXiv:hep-ph/9309216];
B.~P.~Kersevan, M.~Malawski and E.~Richter-Was,
Eur.\ Phys.\ J.\  C {\bf 29} (2003) 541
[arXiv:hep-ph/0207014].


\bibitem{Boos:2010pu}
E.~E.~Boos, S.~V.~Demidov and D.~S.~Gorbunov,
arXiv:1010.5373 [hep-ph].


\bibitem{Eboli:2000ze}
O.~J.~P.~Eboli and D.~Zeppenfeld,
Phys.\ Lett.\  B {\bf 495} (2000) 147
[arXiv:hep-ph/0009158].


\bibitem{King:2005jy}
  S.~F.~King, S.~Moretti and R.~Nevzorov,
  Phys.\ Rev.\  D {\bf 73} (2006) 035009
  [arXiv:hep-ph/0510419].



\bibitem{King:2005my}
  S.~F.~King, S.~Moretti and R.~Nevzorov,
  Phys.\ Lett.\  B {\bf 634} (2006) 278
  [arXiv:hep-ph/0511256].


\bibitem{King:2008qb}
S.~F.~King, R.~Luo, D.~J.~.~Miller and R.~Nevzorov,
JHEP {\bf 0812} (2008) 042
[arXiv:0806.0330 [hep-ph]].

\bibitem{Hall:2009aj}
J.~P.~Hall and S.~F.~King,
JHEP {\bf 0908} (2009) 088
[arXiv:0905.2696 [hep-ph]].


\bibitem{201} E. Keith, E. Ma, Phys. Rev. D {\bf 56} (1997) 7155.

\bibitem{King:2007uj}
S.~F.~King, S.~Moretti, R.~Nevzorov,
Phys.\ Lett.\  B {\bf 650} (2007) 57
[arXiv:hep-ph/0701064].

\bibitem{Accomando:2006ga}
S.~F.~King, S.~Moretti, R.~Nevzorov,
arXiv:hep-ph/0601269;
S. Kraml {\it et al.} (eds.), {\it Workshop on CP studies and
non-standard Higgs physics}, CERN--2006--009, hep-ph/0608079;
S.~F.~King, S.~Moretti, R.~Nevzorov,
AIP Conf.\ Proc.\  {\bf 881} (2007) 138;
R.~Howl, S.~F.~King,
JHEP {\bf 0801} (2008) 030;
S.~F.~King, S.~Moretti, R.~Nevzorov,
{\it  In *Moscow 2006, ICHEP* 1125-1128};
P.~Athron, J.~P.~Hall, R.~Howl, S.~F.~King, D.~J.~Miller, S.~Moretti and R.~Nevzorov,
Nucl.\ Phys.\ Proc.\ Suppl.\  {\bf 200-202} (2010) 120.


\bibitem{202}
  P.~Athron, S.~F.~King, D.~J.~Miller, S.~Moretti and R.~Nevzorov,
  Phys.\ Rev.\  D {\bf 80} (2009) 035009
  [arXiv:0904.2169 [hep-ph]];
  P.~Athron, S.~F.~King, D.~J.~.~Miller, S.~Moretti, R.~Nevzorov and R.~Nevzorov,
  Phys.\ Lett.\  B {\bf 681} (2009) 448
  [arXiv:0901.1192 [hep-ph]];
  P.~Athron, S.~F.~King, D.~J.~Miller, S.~Moretti, R.~Nevzorov,
arXiv:0810.0617 [hep-ph].

\bibitem{203} J. Rich, M. Spiro, J. Lloyd-Owen, Phys. Rept. {\bf 151} (1987) 239;
P.F. Smith, Contemp. Phys. {\bf 29} (1988) 159; T.K. Hemmick {\it et al.},
Phys. Rev. D {\bf 41} (1990) 2074.

\bibitem{Hesselbach:2007te}
S.~Hesselbach, D.~J.~.~Miller, G.~Moortgat-Pick, R.~Nevzorov and M.~Trusov,
Phys.\ Lett.\  B {\bf 662} (2008) 199
[arXiv:0712.2001 [hep-ph]];
S.~Hesselbach, D.~J.~.~Miller, G.~Moortgat-Pick, R.~Nevzorov and M.~Trusov,
arXiv:0710.2550 [hep-ph].


\bibitem{Howl:2009ds}
R.~Howl and S.~F.~King,
Phys.\ Lett.\  B {\bf 687} (2010) 355
[arXiv:0908.2067 [hep-ph]].

\bibitem{Ma:2006ip}
  I.~de Medeiros Varzielas, S.~F.~King and G.~G.~Ross,
  Phys.\ Lett.\  B {\bf 648} (2007) 201
  [arXiv:hep-ph/0607045];
E.~Ma,
Mod.\ Phys.\ Lett.\  A {\bf 21} (2006) 1917
[arXiv:hep-ph/0607056].


\bibitem{adm}
S.~Nussinov,
Phys.\ Lett.\  B {\bf 165} (1985) 55;
S.~M.~Barr, R.~S.~Chivukula and E.~Farhi,
Phys.\ Lett.\  B {\bf 241} (1990) 387;
D.~B.~Kaplan,
Phys.\ Rev.\ Lett.\  {\bf 68} (1992) 741;
D.~Hooper, J.~March-Russell and S.~M.~West,
Phys.\ Lett.\  B {\bf 605} (2005) 228
[arXiv:hep-ph/0410114];
S.~B.~Gudnason, C.~Kouvaris and F.~Sannino,
Phys.\ Rev.\  D {\bf 73} (2006) 115003
[arXiv:hep-ph/0603014];
T.~A.~Ryttov and F.~Sannino,
Phys.\ Rev.\  D {\bf 78} (2008) 115010
[arXiv:0809.0713 [hep-ph]];
D.~E.~Kaplan, M.~A.~Luty and K.~M.~Zurek,
Phys.\ Rev.\  D {\bf 79} (2009) 115016
[arXiv:0901.4117 [hep-ph]];
G.~D.~Kribs, T.~S.~Roy, J.~Terning and K.~M.~Zurek,
Phys.\ Rev.\  D {\bf 81} (2010) 095001
[arXiv:0909.2034 [hep-ph]];
H.~An, S.~L.~Chen, R.~N.~Mohapatra and Y.~Zhang,
JHEP {\bf 1003} (2010) 124
[arXiv:0911.4463 [hep-ph]].

\bibitem{Griest:1986yu}
K.~Griest and D.~Seckel,
Nucl.\ Phys.\  B {\bf 283} (1987) 681
[Erratum-ibid.\  B {\bf 296} (1988) 1034].

\bibitem{adm1}
M.~T.~Frandsen and S.~Sarkar,
Phys.\ Rev.\ Lett.\  {\bf 105} (2010) 011301
[arXiv:1003.4505 [hep-ph]].

\bibitem{Nevzorov:2001um}
P.~A.~Kovalenko, R.~B.~Nevzorov and K.~A.~Ter-Martirosian,
Phys.\ Atom.\ Nucl.\  {\bf 61} (1998) 812
[Yad.\ Fiz.\  {\bf 61} (1998) 898];
R.~B.~Nevzorov, K.~A.~Ter-Martirosyan and M.~A.~Trusov,
Phys.\ Atom.\ Nucl.\  {\bf 65} (2002) 285
[Yad.\ Fiz.\  {\bf 65} (2002) 311]
[arXiv:hep-ph/0105178].

\bibitem{Nevzorov:2000uv}
R.~B.~Nevzorov and M.~A.~Trusov,
J.\ Exp.\ Theor.\ Phys.\  {\bf 91} (2000) 1079
[Zh.\ Eksp.\ Teor.\ Fiz.\  {\bf 91} (2000) 1251]
[arXiv:hep-ph/0106351];

\bibitem{Nevzorov:2004ge}
D.~J.~.~Miller, R.~Nevzorov and P.~M.~Zerwas,
Nucl.\ Phys.\  B {\bf 681} (2004) 3
[arXiv:hep-ph/0304049];
R. Nevzorov, D. J.  Miller, {\it Proceedings to the 7th Workshop "What comes beyond
the Standard Model"}, ed.  by N. S. Mankoc-Borstnik, H. B. Nielsen, C. D. Froggatt,
D. Lukman, DMFA--Zaloznistvo, Ljubljana, 2004, p. 107; hep-ph/0411275.

\bibitem{Miller:2005qua}
C. Panagiotakopoulos, A. Pilaftsis, Phys. Rev. D {\bf 63} (2001) 055003;
D. J. Miller, S. Moretti, R. Nevzorov, {\it Proceedings to the 18th International
Workshop on High-Energy Physics and Quantum Field Theory (QFTHEP 2004)},
ed. by M.N. Dubinin, V.I. Savrin, Moscow, Moscow State Univ., 2004. p. 212; hep-ph/0501139

\bibitem{Djouadi:2005gj}
A.~Djouadi,
Phys.\ Rept.\  {\bf 459} (2008) 1
[arXiv:hep-ph/0503173].

\bibitem{Gorishnii:1990zu}
S.~G.~Gorishnii, A.~L.~Kataev, S.~A.~Larin and L.~R.~Surguladze,
Mod.\ Phys.\ Lett.\  A {\bf 5} (1990) 2703;
S.~G.~Gorishnii, A.~L.~Kataev, S.~A.~Larin and L.~R.~Surguladze,
Phys.\ Rev.\  D {\bf 43} (1991) 1633;
K.~G.~Chetyrkin and A.~Kwiatkowski,
Nucl.\ Phys.\  B {\bf 461} (1996) 3
[arXiv:hep-ph/9505358];
S.~A.~Larin, T.~van Ritbergen and J.~A.~M.~Vermaseren,
Phys.\ Lett.\  B {\bf 362} (1995) 134
[arXiv:hep-ph/9506465].

\bibitem{Heinemeyer:2004ms}
S.~Heinemeyer,
Int.\ J.\ Mod.\ Phys.\  A {\bf 21} (2006) 2659
[arXiv:hep-ph/0407244].

\bibitem{cdm}
J.~Dunkley {\it et al.}  [WMAP Collaboration],
Astrophys.\ J.\ Suppl.\  {\bf 180} (2009) 306
[arXiv:0803.0586 [astro-ph]].

\bibitem{Wells:1997ag}
J.~D.~Wells,
arXiv:hep-ph/9708285.

\bibitem{Barger:2004bz}
V.~Barger, C.~Kao, P.~Langacker and H.~S.~Lee,
Phys.\ Lett.\  B {\bf 600} (2004) 104
[arXiv:hep-ph/0408120];
A.~Menon, D.~E.~Morrissey and C.~E.~M.~Wagner,
Phys.\ Rev.\  D {\bf 70} (2004) 035005
[arXiv:hep-ph/0404184];
V.~Barger, P.~Langacker and H.~S.~Lee,
Phys.\ Lett.\  B {\bf 630} (2005) 85
[arXiv:hep-ph/0508027];
S.~V.~Demidov and D.~S.~Gorbunov,
JHEP {\bf 0702} (2007) 055
[arXiv:hep-ph/0612368];
V.~Barger, P.~Langacker, I.~Lewis, M.~McCaskey, G.~Shaughnessy and B.~Yencho,
Phys.\ Rev.\  D {\bf 75} (2007) 115002
[arXiv:hep-ph/0702036].

\bibitem{MicrOMEGA}
G.~Belanger, F.~Boudjema, A.~Pukhov and A.~Semenov,
arXiv:0803.2360 [hep-ph];
G.~Belanger, F.~Boudjema, A.~Pukhov and A.~Semenov,
Comput.\ Phys.\ Commun.\  {\bf 176}, 367 (2007)
[arXiv:hep-ph/0607059];
G.~Belanger, F.~Boudjema, A.~Pukhov and A.~Semenov,
Comput.\ Phys.\ Commun.\  {\bf 174} (2006) 577
[arXiv:hep-ph/0405253];
G.~Belanger, F.~Boudjema, A.~Pukhov and A.~Semenov,
arXiv:hep-ph/0112278.

\bibitem{LanHEP}
A.~Semenov,
arXiv:0805.0555 [hep-ph];
A.~V.~Semenov,
arXiv:hep-ph/0208011.


\bibitem{Erler:2010uy}
J.~Erler, P.~Langacker, S.~Munir and E.~rojas,
arXiv:1010.3097 [hep-ph];
C.Hays, A.Kotwal, O.Stelzer-Chilton, ``A Search for Dimuon
Resonances with CDF in Run II'',
{\tt
http://www-cdf.fnal.gov/physics/exotic/r2a/20080710.dimuon\_resonance/}

\bibitem{Erler:2009jh}
J.~Erler, P.~Langacker, S.~Munir and E.~R.~Pena,
JHEP {\bf 0908} (2009) 017
[arXiv:0906.2435 [hep-ph]].

\bibitem{Baer:1992dc}
H.~Baer and X.~Tata,
Phys.\ Rev.\  D {\bf 47} (1993) 2739;
H.~Baer, C.~Kao and X.~Tata,
Phys.\ Rev.\  D {\bf 48} (1993) 5175
[arXiv:hep-ph/9307347].


\bibitem{trilepton}
O.~Mundal  [D0 Collaboration],
arXiv:0710.4098 [hep-ex];
T.~Aaltonen {\it et al.}  [CDF Collaboration],
Phys.\ Rev.\ Lett.\  {\bf 101} (2008) 251801
[arXiv:0808.2446 [hep-ex]];
V.~M.~Abazov {\it et al.}  [D0 Collaboration],
Phys.\ Lett.\  B {\bf 680} (2009) 34
[arXiv:0901.0646 [hep-ex]];
J.~Strologas  [CDF Collaboration],
AIP Conf.\ Proc.\  {\bf 1200} (2010) 275
[arXiv:0910.1889 [hep-ex]].

\bibitem{Kraan:2005vy}
A.~C.~Kraan,
arXiv:hep-ex/0505002.

\bibitem{pdg}
[ALEPH Collaboration and DELPHI Collaboration and L3 Collaboration],
Phys.\ Rept.\ {\bf 427} (2006) 257 [arXiv:hep-ex/0509008];
C.~Amsler {\it et al.} [Particle Data Group],
Phys.\ Lett.\  B {\bf 667} (2008) 1.

\bibitem{Z-inv-width}
J.~Abdallah {\it et al.} [DELPHI Collaboration],
Eur.\ Phys.\ J.\ C {\bf 38} (2005) 395
[arXiv:hep-ex/0406019].

\bibitem{Abbiendi:2003sc}
G.~Abbiendi {\it et al.} [OPAL Collaboration],
Eur.\ Phys.\ J.\ C {\bf 35} (2004) 1
[arXiv:hep-ex/0401026].


\bibitem{Aprile:2010um}
E.~Aprile {\it et al.}  [XENON100 Collaboration],
Phys.\ Rev.\ Lett.\  {\bf 105} (2010) 131302
[arXiv:1005.0380 [astro-ph.CO]].

\bibitem{Ellis:2008hf}
J.~R.~Ellis, K.~A.~Olive and C.~Savage,
Phys.\ Rev.\  D {\bf 77} (2008) 065026
[arXiv:0801.3656 [hep-ph]].

\bibitem{Kalinowski:2008iq}
J.~Kalinowski, S.~F.~King and J.~P.~Roberts,
JHEP {\bf 0901} (2009) 066
[arXiv:0811.2204 [hep-ph]].

\bibitem{TuckerSmith:2001hy}
D.~Tucker-Smith and N.~Weiner,
Phys.\ Rev.\  D {\bf 64} (2001) 043502
[arXiv:hep-ph/0101138].

\bibitem{Cheung:2009wb}
K.~Cheung and T.~C.~Yuan,
Phys.\ Lett.\  B {\bf 685} (2010) 182
[arXiv:0912.4599 [hep-ph]];
K.~Cheung, K.~H.~Tsao and T.~C.~Yuan,
arXiv:1003.4611 [hep-ph].

\bibitem{Bottino:1999ei}
A.~Bottino, F.~Donato, N.~Fornengo and S.~Scopel,
Astropart.\ Phys.\  {\bf 13} (2000) 215
[arXiv:hep-ph/9909228];
J.~R.~Ellis, A.~Ferstl and K.~A.~Olive,
Phys.\ Lett.\  B {\bf 481} (2000) 304
[arXiv:hep-ph/0001005];
J.~R.~Ellis, A.~Ferstl and K.~A.~Olive,
Phys.\ Rev.\  D {\bf 63} (2001) 065016
[arXiv:hep-ph/0007113];
A.~Bottino, F.~Donato, N.~Fornengo and S.~Scopel,
Astropart.\ Phys.\  {\bf 18} (2002) 205
[arXiv:hep-ph/0111229].

\bibitem{Collar:2010ht}
J.~I.~Collar,
arXiv:1010.5187 [astro-ph.IM].


\bibitem{King:2010np}
  S.~F.~King,
  JHEP {\bf 1009} (2010) 114
  [arXiv:1006.5895 [hep-ph]].

\end{thebibliography}
\end{document}